# Decentralized Maximum Likelihood Estimation for Sensor Networks Composed of Nonlinearly Coupled Dynamical Systems

Sergio Barbarossa, Gesualdo Scutari*

Dpt. INFOCOM, Univ. of Rome "La Sapienza ", Via Eudossiana 18, 00184 Rome, Italy

E-mail: {`sergio, aldo.scutari`}@infocom.uniroma1.it.

**Abstract**

In this paper we propose a decentralized sensor network scheme capable to reach a globally optimum maximum likelihood (ML) estimate through self-synchronization of nonlinearly coupled dynamical systems. Each node of the network is composed of a sensor and a first-order dynamical system initialized with the local measurements. Nearby nodes interact with each other exchanging their state value and the final estimate is associated to the state derivative of each dynamical system. We derive the conditions on the coupling mechanism guaranteeing that, if the network observes one common phenomenon, each node converges to the globally optimal ML estimate. We prove that the synchronized state is globally asymptotically stable if the coupling strength exceeds a given threshold. Acting on a single parameter, the coupling strength, we show how, in the case of nonlinear coupling, the network behavior can switch from a global consensus system to a spatial clustering system. Finally, we show the effect of the network topology on the scalability properties of the network and we validate our theoretical findings with simulation results.

## 1  Introduction

Sensor networks are receiving a significant attention because of their many potential civilian and military applications (see, e.g., [1, 2, 3]). The single major challenge is perhaps how to conjugate the relative unreliability of the single node, due to its limited complexity and energy availability, with the high reliability required to the whole network. Most research works aim then at making the best use of the

*This work has been partially funded by ARL, Contract N62558-05-P-0458, and by the WINSOC project, a Specific Targeted Research Project (Contract Number 0033914) co-funded by the INFSO DG of the European Commission within the RTD activities of the Thematic Priority Information Society Technologies.

available resources. Many works concentrate on how to adapt the protocol stacks derived in decades of research in communication networks to the sensor network scenario. An alternative approach consists, instead, in recognizing that a sensor network is intrinsically different from a communication network, thus implying that the design of a sensor network should reflect its specificities. Among the features distinguishing a sensor network from a communication network, we may mention its *data-centric* and *event-driven* nature. In general, a sensor network can be seen as a sort of distributed computer that, on the basis of the measurements, let us say $x_1, x_2, \ldots, x_N$, gathered by $N$ sensors, it has to take a decision about the observed phenomenon by computing a function $f(x_1, x_2, \ldots, x_N)$ of the measurements. Typically, this function has properties, depending on the application, that, if properly exploited, can suggest efficient ways to design a sensor network. For example, Giridhar and Kumar recently proved that, if $f(x_1, x_2, \ldots, x_N)$ is invariant to any permutation of the observed variables (like in the computation of the average, for example, or the maximum, etc.), it is possible to improve the scalability properties of the network, using some kind of *in-network* processing [4]. Interestingly, this symmetry property is not at all artificial, as it reflects the data-centric nature of the network and it holds true in a variety of applications. Some works exploit the data-centric property to devise innovative schemes, like, for example, the type-based multiple access (TBMA) system [6]. TBMA is perfectly scalable, but it requires a high coherence of the channels from the sensors to the sink node.

Quite recently, several authors have proposed an alternative approach that allows each node to perform in-network processing, so as to reduce the burden of the fusion center [1]. Other works go even further by proposing strategies where the global decision, or estimation, is obtained using a totally distributed approach, *with no need for a fusion center*, at least in the case where the whole network observes a common event [7, 8, 9, 10, 11, 12, 19]. A strategy that has received significant attention in the last few years is the so called *average consensus* protocol. The basic idea is that, if the network is connected, i.e., there is a path, possibly composed of multiple hops, between any pair of nodes, *local* exchange of information among nearby sensors is sufficient to reach a *global* consensus on the average of the observed values, *without requiring any control node*. Global consensus can be reached through linear coupling, as in [19], [9], or through nonlinear coupling, as in [13], [14], [34]. Global consensus can also be used to track a common time-varying phenomenon, as in [17], [16]. An important synergism to this approach comes also from the algorithms developed for the coordination of groups of mobile autonomous agents through local transmissions [21]. An alternative approach to achieve a consensus was proposed in [22, 23, 24], where consensus was seen as a result of self-synchronization of a population of pulse-coupled oscillators, each one initialized with the sensor local estimates or decisions. The principle ensuring the self-synchronization capability of the system proposed in [22], [23] relied on a theorem, proved by Mirollo and Strogatz in [27], that required the *full* network connectivity, i.e., the property that each node has a direct link to each other node. This assumption was later removed by Lucarelli and Wang in [24], who proved that local coupling among the nodes is sufficient, provided that the whole network is connected. The pulse coupling mechanism is indeed appealing from the implementation point of view, but, especially for large scale networks, it may suffer from ambiguity problems, as the information bearing time shift may become indistinguishable from the propagation delay. The idea of achieving global estimates or decisions exploiting local coupling among dynamical systems, initialized with local measurements, was then proposed in [12], [25], [26]. Average consensus through mutual coupling of first order dynamical systems was used in [12], [26] to derive the globally optimal maximum likelihood estimation: In [12], each system is initialized with the local measurement, the coupling is linear and the consensus amounts to requiring all dynamical systems to reach the same value of the state; in [26], each system is randomly initialized, the coupling is nonlinear (it subsumes linear coupling as a particular case) and the consensus refers to the situation where the derivatives of the states (rather than the states) converge, asymptotically, to a common value.

This paper builds on the initial idea of [26] and its main contributions are the following: i) we derive the conditions guaranteeing that a globally optimum maximum likelihood estimator can be obtained through local nonlinear coupling of first order dynamical systems; ii) we show that nonlinear coupling offers a variety of behaviors, to be used to find out the best implementation of the radio transceivers or to allow the network to work as a global estimator or as a spatial clustering mechanism; iii) we show that convergence on the state derivative (rater than the state) improves the resilience against additive noise with respect to common average consensus techniques.

The paper is organized as follows. In Section 2 we describe the coupling mechanism. In Section 3 we show how to design the coupling mechanism and the local processing in order to make the equilibrium achievable by each node to coincide with the globally optimum maximum likelihood estimate. In Section 4 we derive the conditions guaranteeing that the equilibrium is unique and asymptotically stable. Finally, in Section 5 we report numerical results validating our theoretical findings and showing the network behavior both as a global estimator or as a spatial clustering system.

## 2  Coupling mechanism

The proposed sensor network is composed of $N$ nodes, each composed of four basic components: i) a *transducer* that senses the physical parameter of interest (e.g., temperature, concentration of contaminants, radiation, etc.); ii) a *local detector* or *estimator* that processes the measurements taken by the node; iii) a *dynamical system* whose state evolves according to a first-order differential equation, whose parameters depend on the local estimate and on the states of nearby nodes; iv) a *radio interface* that transmits the state of the dynamical system and receives the state transmitted by nearby nodes, thus ensuring the interaction among nearby nodes.

## 2.1 Scalar observations

When each sensor measures a single physical parameter, the dynamical system present in node $i$ evolves according to the following equation

$$\dot{\theta}_i(t) = \omega_i + \frac{K}{c_i} \sum_{j=1}^{N} a_{ij}\, f\left[\theta_j(t) - \theta_i(t)\right] + v_i(t), \qquad i = 1, \ldots, N, \tag{1}$$

$$\theta_i(0) = \theta_{i0},$$

where

1. $\theta_i(t)$ is the state function of the $i$-th sensor, initialized, at $t=0$, as any random number $\theta_i(0) = \theta_{i0}$;

2. $\omega_i$ is a function $g(x_i)$ of the observation $x_i$ taken from node $i$;

3. $f(\cdot)$ is a nonlinear, odd function that takes into account the coupling among the sensors[1];

4. $K$ is a positive control loop gain measuring the coupling strength;

5. $c_i$ is a positive coefficient that quantifies the attitude of the $i$-th sensor to adapt its state as a function of the signals received from the other nodes;

6. the coefficients $a_{ij}$ take into account the local coupling among the systems: if nodes $i$ and $j$ are coupled to each other, $a_{ij} \neq 0$, otherwise $a_{ij} = 0$; we assume that the nonzero coefficients $a_{ij}$ are positive and respect the symmetry condition $a_{ij} = a_{ji}$;

7. $v_i(t)$ is additive noise.

The model (1) coincides with the so called Kuramoto model [28], when $f(x) = \sin(x)$ and $a_{ij} = c_i = 1, \forall i,j$ [2]. Given the model (1), the running decision, or estimate, of each sensor is associated to the *derivative* of the state function $\dot{\theta}_i(t)$. Global consensus, in this paper, means that all nodes end up evolving with the same *state derivative*. This choice is different from common average consensus techniques, where the consensus refers to the state value. We will show, in Section 5, that this apparently slight difference brings important consequences in the presence of additive coupling noise. Furthermore, in our case, the states of different nodes are let free to converge to functions differing by a constant term. This extra degree of freedom, with respect to the techniques converging on the state, might be exploited for different scopes than synchronization, like, for example, spatial pattern recognition, as in [30].

A possible schematic implementation of our protocol is reported in Fig. 1. On the right side, there is a transducer that measures a physical quantity $x_i$ and produces a parameter $\omega_i = g(x_i)$. On the left side, there is a single transmit/receive antenna and a circulator used to switch between transmission and reception. The transmitted signal is a waveform $p(t; \theta_i(t))$ that depends on the local state $\theta_i(t)$. The received signal is a linear combination of the signals transmitted by nearby nodes, i.e.

---

[1] We assume, w.l.o.g., that $f(x)$ is normalized so that $f'(0) = 1$, where $f'(x) := df(x)/dx$, as different values of $f'(0)$ can always be included in $K$.

[2] As will be clarified in Section 4, our function $f(x)$ has to be a monotonically increasing function, to guarantee the achievement of the global optimal ML estimate. Hence, Kuramoto model is mentioned here only for similarity reasons, but our main findings do not include Kuramoto model as a particular case.

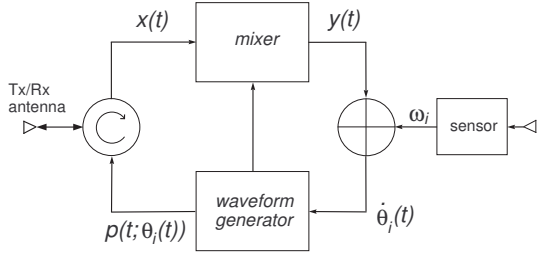

Figure 1: General coupling mechanism.

$$x(t) = \sum_{j=1}^{N} a_{ij}\, p(t; \theta_j(t)),$$

where the coefficients $a_{ij}$ depends on the propagation and radio interface. The received signal is then mixed with a local waveform $q(t; \theta_i(t))$, generated as a function of $\dot{\theta}_i(t)$, in order to produce the signal

$$y(t) = \sum_{j=1}^{N} a_{ij}\, f[\theta_j(t) - \theta_i(t)].$$

It is easy to check that the input of the waveform generator coincides with (1).

The scheme of Fig. 1 is rather general and it incorporates alternative implementations, like pulse coupled systems, for example, or phase-locked loops, depending on the choice of the waveforms $p(t)$ and $q(t)$. The half-duplex feature of the scheme depicted in Fig. 1 implies that, if two nodes $i$ and $j$ transmit the same waveform, i.e. $\theta_i(t) = \theta_j(t)$, they do not listen to each other. This feature is consistent, in mathematical terms, with the odd property of the function $f(x)$, as will be described in Section 4.

We assume, initially, that the additive noise is negligible, i.e. $v_i(t) = 0$. In Section 5, we will show the effect of noise on the system performance.

To make explicit the network connectivity properties, it is useful to rewrite (1) introducing the graph *incidence* matrix $\boldsymbol{B}$, defined as follows. Given an oriented graph $\mathscr{G}^3$ composed by $N$ vertices and $E$ edges, $\boldsymbol{B}$ is the $N \times E$ matrix such that $[\boldsymbol{B}]_{ij} = 1$ if the edge $j$ is incoming to vertex $i$, $[\boldsymbol{B}]_{ij} = -1$ if the edge $j$ is outcoming from vertex $i$, and 0 otherwise. Given the $N \times 1$ vector $\mathbf{1}_N$, composed of all ones, it is easy to check that the incidence matrix satisfies the following property:

$$\mathbf{1}_N^T \boldsymbol{B} = \mathbf{0}_E^T. \tag{2}$$

Given $\boldsymbol{B}$, the symmetric $N \times N$ matrix $\boldsymbol{L}$ defined as $\boldsymbol{L} \triangleq \boldsymbol{B}\boldsymbol{B}^T$, is called the *Laplacian* of $\mathscr{G}$ and it is independent of the graph orientation. If we associate a positive number $w_i$ to each edge and we build the diagonal matrix $\boldsymbol{D_w} \triangleq \operatorname{diag}(\mathbf{w})$, with $\mathbf{w} \triangleq [w_1, \cdots, w_E]^T$, we may introduce the so called *weighted Laplacian*, which is written as $\mathbf{L_w} \triangleq \boldsymbol{B} \boldsymbol{D_w} \boldsymbol{B}^T$. The Laplacian (as well as the weighted Laplacian) has several important properties, among which [31]: 1) $\boldsymbol{L}$ (or $\mathbf{L_w}$) is always positive semi-definite, i.e. with the smallest eigenvalue always equal to zero; 2) the algebraic multiplicity of the null eigenvalue is equal to the number $n_c$ of connected components of the graph; if the graph is connected, $n_c = 1$ and $\operatorname{rank}(\boldsymbol{L}) = \operatorname{rank}(\mathbf{L_w}) = N - 1$, i.e. $\boldsymbol{L}$ (or $\mathbf{L_w}$) has a unique zero eigenvalue and the eigenvector associated to the null eigenvalue is the vector $\mathbf{1}_N$. The second smallest eigenvalue $\lambda_2(\boldsymbol{L})$ (or $\lambda_2(\mathbf{L_w})$) is known as the graph *algebraic connectivity*, and it provides a measure of connectivity [32].

---
[3]The orientation of a graph $\mathscr{G}$ consists in the assignment of a direction to each edge.

Using the above notation, $\mathbf{L_A} \triangleq \boldsymbol{B} \boldsymbol{D_A} \boldsymbol{B}^T$ will denote the weighted Laplacian associated to the graph describing our network (1), including the positive coefficients $\{a_{ij}\}$. Furthermore, $d_{\max} \triangleq \max_i \sum_{j=1}^N a_{ij}$ will denote the maximum degree of the (weighted) graph.

Using the incidence matrix $\boldsymbol{B}$, we can rewrite (1) in compact form as

$$\dot{\boldsymbol{\theta}}(t) = \boldsymbol{\omega} - K \boldsymbol{D_c}^{-1} \boldsymbol{B} \boldsymbol{D_A} f\left[\boldsymbol{B}^T \boldsymbol{\theta}(t)\right], \qquad (3)$$
$$\boldsymbol{\theta}(0) = \boldsymbol{\theta}_0,$$

where $\boldsymbol{\theta}(t) \triangleq [\theta_1(t), \cdots, \theta_N(t)]^T$; $\boldsymbol{\theta}_0 \triangleq [\theta_{10}, \cdots, \theta_{N0}]^T$; $\boldsymbol{D_c} \triangleq \operatorname{diag}\{c_1, \ldots, c_N\}$; $\boldsymbol{D_A}$ is an $E \times E$ diagonal matrix, whose diagonal entries are all the weights $a_{ij}$, indexed from 1 to $E$; the symbol $f(\boldsymbol{x})$ has to be intended as the vector whose $k$-th component is $f(x_k)$.

## 2.2 Vector Observation

When each sensor measures more, let us say $L$, physical parameters like, e.g., temperature, pressure, etc., the coupling mechanism (1) generalizes according to the following expression[4]

$$\dot{\boldsymbol{\theta}}_i(t) = \boldsymbol{\omega}_i + K \mathbf{Q}_i^{-1} \sum_{j=1}^N a_{ij} f\left[\boldsymbol{\theta}_j(t) - \boldsymbol{\theta}_i(t)\right], \qquad i = 1, \ldots, N, \qquad (4)$$
$$\boldsymbol{\theta}_i(0) = \boldsymbol{\theta}_{i0},$$

where $\boldsymbol{\theta}_i(t)$ is the $L$-size vector state of the $i$-th node, that is initialized as a random vector $\boldsymbol{\theta}_{i0}$; $\boldsymbol{\omega}_i$ is the $L$-size vector, function of the $L$ measurements taken by node $i$; $\mathbf{Q}_i$ is an $L \times L$ non-singular matrix that depends on the observation model. In Section 3, we show how to choose the vectors $\boldsymbol{\omega}_i$ and the matrices $\mathbf{Q}_i$ to guarantee the convergence of (4) to the global optimal maximum likelihood (ML) estimate. Also in this case, we can rewrite (4) in compact form using the graph incidence matrix. Introducing the vectors $\boldsymbol{\theta}(t) \triangleq [\boldsymbol{\theta}_1^T(t), \ldots \boldsymbol{\theta}_N^T(t)]^T$ and $\boldsymbol{\omega} \triangleq [\boldsymbol{\omega}_1^T, \ldots, \boldsymbol{\omega}_N^T]^T$, and the matrix $\mathbf{D_Q} \triangleq \operatorname{diag}(\boldsymbol{Q}_1, \ldots, \boldsymbol{Q}_N)$, the system (4) becomes

$$\dot{\boldsymbol{\theta}}(t) = \boldsymbol{\omega} - K \mathbf{D_Q}^{-1} \mathbf{P}^T (\boldsymbol{I}_L \otimes \boldsymbol{B} \boldsymbol{D_A}) f\left[(\boldsymbol{I}_L \otimes \boldsymbol{B}^T) \mathbf{P} \boldsymbol{\theta}(t)\right], \qquad (5)$$
$$\boldsymbol{\theta}(0) = \boldsymbol{\theta}_0,$$

---
[4] We assume that the coupling coefficients $a_{ij}$ are the same for all estimated parameters. This assumption is justified by the fact that $a_{ij}$ depends on the coverage radius of each transmitter and not on the measurements.

where $\otimes$ denotes the Kronecker product, $\boldsymbol{\theta}(0) \triangleq [\boldsymbol{\theta}_1^T(0), \ldots \boldsymbol{\theta}_N^T(0)]^T$ and $\mathbf{P}$ is an $LN \times LN$ permutation matrix defined as

$$[\mathbf{P}]_{ij} = \begin{cases} 1, & \text{if } j = (((i \bmod N) - 1)L + \left\lceil \dfrac{i}{N} \right\rceil \bmod(NL), \\ 0, & \text{otherwise}, \end{cases} \qquad (6)$$

so that, in (5), each vector $\mathbf{x} = [\mathbf{x}_1^T, \ldots, \mathbf{x}_N^T]^T$, with $\mathbf{x}_i = [x_i^{(1)}, \ldots, x_i^{(L)}]^T$, is mapped into a new vector $\bar{\mathbf{x}} = \mathbf{P}\mathbf{x}$, partitioned as $\bar{\mathbf{x}} = [\bar{\mathbf{x}}_1^T, \ldots, \bar{\mathbf{x}}_L^T]^T$, with $\bar{\mathbf{x}}_i = [x_1^{(i)}, \ldots, x_N^{(i)}]^T$.

## 2.3 Network Self-Synchronization

Differently from [12], where the global consensus was intended to be the situation where all dynamical systems reach the same state, we adopt an alternative consensus strategy. We define the network synchronization with respect to the state derivative, as follows:

**Definition 1** *The overall population of dynamical systems (1) (or (4)) is said to synchronize if there exists a solution $\boldsymbol{\theta}^\star(t)$ of (1) (or (4)) such that all the state derivatives converge, asymptotically, to $\dot{\boldsymbol{\theta}}^\star(t)$, i.e.*

$$\lim_{t \mapsto \infty} \|\dot{\boldsymbol{\theta}}_i(t) - \dot{\boldsymbol{\theta}}^\star(t)\| = 0, \quad i = 1, 2, \ldots, N, \qquad (7)$$

*where $\|\cdot\|$ denotes some vector norm. The system is* globally asymptotically stable *if the system synchronizes, in the sense specified before, for* any *set of initial conditions $\boldsymbol{\theta}_i(0)$.*

According to Definition 1, if there exists a synchronized state that is globally asymptotically stable, then it must necessarily be *unique*. Interestingly, if the synchronized state exists, it can be computed in closed form, without explicitly solving the system of differential equations (1) and (4). In fact, exploiting the oddness property of $f(x)$, left-multiplying (3) by the row vector $\mathbf{c}^T \triangleq \mathbf{1}_N^T \mathbf{D}_\mathbf{c}$, we obtain

$$\mathbf{c}^T \dot{\boldsymbol{\theta}}(t) = \mathbf{c}^T \boldsymbol{\omega} - K \mathbf{1}_N^T \mathbf{B} \mathbf{D}_\mathbf{A} f\left(\mathbf{B}^T \boldsymbol{\theta}\right) = \mathbf{c}^T \boldsymbol{\omega}, \qquad (8)$$

where in the second equality of (8), we have used (2). Hence, if system (3) synchronizes (according to Definition 1), the common value of $\dot{\boldsymbol{\theta}}^\star(t)$ must be constant and equal to

$$\dot{\boldsymbol{\theta}}^\star(t) \triangleq \omega^\star = \frac{\mathbf{c}^T \boldsymbol{\omega}}{\mathbf{1}_N^T \mathbf{c}} = \frac{\sum_{i=1}^N c_i \omega_i}{\sum_{i=1}^N c_i}. \qquad (9)$$

Similarly, in the vector case, left-multiplying (5) by the matrix $\left(\mathbf{1}_N^T \otimes \mathbf{I}_L\right) \mathbf{D}_Q$, we obtain

$$\sum_{i=1}^N \mathbf{Q}_i \dot{\boldsymbol{\theta}}_i(t) = \sum_{i=1}^N \mathbf{Q}_i \boldsymbol{\omega}_i, \qquad (10)$$

where we used the following chain of equalities $(\mathbf{1}_N^T \otimes \mathbf{I}_L)\mathbf{P}^T(\mathbf{I}_L \otimes \mathbf{B} \mathbf{D}_\mathbf{A}) = (\mathbf{I}_L \otimes \mathbf{1}_N^T)(\mathbf{I}_L \otimes \mathbf{B} \mathbf{D}_\mathbf{A}) = \mathbf{I}_L \otimes \mathbf{1}_N^T \mathbf{B} \mathbf{D}_\mathbf{A} = \mathbf{0}_{L \times LE}$, and the property (2). Hence, if the system synchronizes, in the sense of Definition

1, the synchronized state must necessarily be

$$\dot{\boldsymbol{\theta}}^\star(t) \triangleq \boldsymbol{\omega}_L^\star = \left(\sum_{i=1}^N \mathbf{Q}_i\right)^{-1} \left(\sum_{i=1}^N \mathbf{Q}_i \boldsymbol{\omega}_i\right). \tag{11}$$

## 3 Reaching global ML estimate through self-synchronization

The basic idea of this paper is that, when the whole network observes a common phenomenon, the self-synchronization process forms the basic mechanism for reaching the globally optimal ML estimate through local exchange of the state functions, *without sending the observations to any fusion center*. In particular, let us consider the scalar observation

$$x_i = b_i \xi + w_i, \tag{12}$$

where $\xi$ is the common unknown parameter to be estimated and $w_i$, $i = 1, \ldots, N$ are a set of i.i.d. Gaussian random variables with zero mean and variances $(\sigma_1^2, \ldots, \sigma_N^2)$. Initializing each node with $\omega_i = x_i/b_i$ and setting, in (3), $c_i = b_i^2/\sigma_i^2$, the network synchronized state (9) becomes:

$$\omega^\star = \hat{\xi}_{ML} = \frac{\sum_{i=1}^N \frac{b_i x_i}{\sigma_i^2}}{\sum_{i=1}^N \frac{b_i^2}{\sigma_i^2}}. \tag{13}$$

This value coincides with the globally optimal ML estimate [33]. The equilibrium (13) shows that the most reliable nodes (i.e., the ones with the smallest $\sigma_i$) are the most influent nodes in driving the whole system towards the common decision. What is important to stress is that this happens without any node knowing which are the nodes with the best SNR.

In the linear vector case, each node observes the vector

$$\boldsymbol{x}_i = \boldsymbol{A}_i \boldsymbol{\xi} + \boldsymbol{w}_i, \tag{14}$$

where $\boldsymbol{x}_i$ is the $M \times 1$ observation vector, $\boldsymbol{\xi}$ is the $L \times 1$ unknown common parameter vector, $\boldsymbol{A}_i$ is the $M \times L$ mixing matrix, and $\boldsymbol{w}_i$ is the observation noise vector, modeled as a Gaussian vector with zero mean and covariance matrix $\boldsymbol{C}_i$. We assume that the noise vectors affecting different sensors are statistically independent of each other (however, the noise vector present in each sensor may be colored). We consider the case where the single sensor must be able, in principle, to recover the parameter vector $\boldsymbol{\xi}$ from its own observation. This requires that $M \geq L$ and that $\boldsymbol{A}_i$ is full column rank. In this case, setting, in (5), $\boldsymbol{Q}_i = \boldsymbol{A}_i^H \boldsymbol{C}_i^{-1} \boldsymbol{A}_i$ and $\boldsymbol{\omega}_i = \left(\boldsymbol{A}_i^H \boldsymbol{C}_i^{-1} \boldsymbol{A}_i\right)^{-1} \boldsymbol{A}_i^H \boldsymbol{C}_i^{-1} \boldsymbol{x}_i$, the synchronized state (11) becomes

$$\boldsymbol{\omega}_L^\star = \hat{\boldsymbol{\xi}}_{ML} = \left(\sum_{i=1}^n \boldsymbol{A}_i^H \boldsymbol{C}_i^{-1} \boldsymbol{A}_i\right)^{-1} \left(\sum_{i=1}^n \boldsymbol{A}_i^H \boldsymbol{C}_i^{-1} \boldsymbol{x}_i\right), \tag{15}$$

which coincides with the globally optimal ML estimate [33].

It is important to emphasize here that a virtually optimal fusion center in this case would need to know not only the observation vectors $\boldsymbol{x}_i$, but also all mixing matrices $\boldsymbol{A}_i$ and the noise covariance matrices $\boldsymbol{C}_i$. Conversely, following the proposed approach, if the network converges, each node tends to the optimal ML estimate without sending all these data to any sink node, but simply exchanging the state vectors $\boldsymbol{\theta}_i(t)$ with nearby nodes. The penalty for having this advantage is that the solution is reached through an iterative procedure that consumes time and energy. However, this energy is spent only for local transmissions. The crucial point, as far as the global energy consumption is concerned, is then the convergence time. In the next section, we will give an approximate formula for such a value.

The proposed system has indeed a broader applicability than just global ML estimation for a linear observation model. From (9), since $\omega_i$ is a function $g_i(x_i)$ of the local observation, our approach allows us to compute any function of the collected data expressible in the form

$$f(x_1, x_2, \ldots, x_N) = h\left[\frac{\sum_{i=1}^{N} c_i g(x_i)}{\sum_{i=1}^{N} c_i}\right], \qquad (16)$$

with positive coefficients $c_i$, with a totally distributed mechanism. Of course, this class of functions is not the most general one. Nevertheless, the class of functions in (16) contains not only the linear ML case, but it comprises many cases of practical interest (like, for example, computation of the sufficient statistics in detection of Gaussian processes in Gaussian noise, computation of maximum, minimum, histograms, geometric mean of the observed values, etc.), as it can be checked by choosing the functions $g(x), h(x)$ and the coefficients $c_i$ appropriately.

## 4 Global asymptotic stability of the synchronized state

Given the dynamical system (1) or (4), the natural questions to ask are: i) Does the synchronized state exist? ii) If it exists, does the system synchronize for any set of initial conditions? In this section, we provide an answer to these questions. We focus, initially, on the scalar system (1) and provide necessary and sufficient conditions for the existence of a globally asymptotically synchronized state, according to Definition 1. Then, we generalize the result to the vector system (4).

**Theorem 1** *Given the system (1), assume that the following conditions are satisfied:*

**a1** *The graph associated to the network is connected;*

**a2** *The nonlinear function $f(\cdot) : \mathbb{R} \mapsto \mathbb{R}$ is a continuously differentiable, odd, increasing function;*

**a3** *The nonzero coefficients $a_{ij}$ and the coefficients $c_i$ are positive.*

*Then, there exist two unique critical values of $K$, denoted by $K_L$ and $K_U$, with $0 \leq K_L \leq K_U$, such that the synchronized state exists for all $K > K_U$, and it does not for all $K < K_L$. Furthermore, if it* exists, the synchronized state is globally asymptotically stable. Lower and upper bounds of $K_L$ and $K_U$ are given by

$$K_L \geq \frac{\|\boldsymbol{D_c}\Delta\boldsymbol{\omega}\|_\infty}{f_{\max}d_{\max}}, \quad and \quad K_U \leq \frac{\|\boldsymbol{D_c}\Delta\boldsymbol{\omega}\|_2}{g\lambda_2(\boldsymbol{L_A})}, \tag{17}$$

where

$$g = \sup\left\{a \min_{x\in[0,2a]} \frac{f(x)}{x}, \quad a \in \mathbb{R}_{++}\right\} \in (0, f_{\max}/2]; \tag{18}$$

$\Delta\boldsymbol{\omega} \triangleq \boldsymbol{\omega} - \omega^*\boldsymbol{1_N}$, with $\omega^*$ defined in (9); $f_{\max} \triangleq \lim_{x\to+\infty} f(x)^5$; $d_{\max}$ and $\lambda_2(\boldsymbol{L_A})$ are the maximum degree and the algebraic connectivity of the graph, respectively.

**Proof.** See Appendix A. ∎

**Corollary 1** *Assume that $f(\cdot)$, in addition to **a2**, is asymptotically convex or concave[6]. Then:*

1. *If $f(\cdot)$ is unbounded, $K_L = K_U = 0$;*

2. *If $f(\cdot)$ is bounded, i.e., $f_{max} < \infty$, upper and lower bounds of $K_L$ and $K_U$ are*

$$K_L \geq \frac{\|\boldsymbol{D_c}\Delta\boldsymbol{\omega}\|_\infty}{f_{\max}d_{\max}}, \quad and \quad K_U \leq \frac{2\|\boldsymbol{D_c}\Delta\boldsymbol{\omega}\|_2}{f_{\max}\lambda_2(\boldsymbol{L_A})}. \tag{19}$$

**Remark 1.** Even though conditions (17) or (19) provide only a range of values for $K_L$ and $K_U$, they state an important property of the whole system: If we want the network to reach a global consensus (common estimate), it is sufficient to take $K$ greater than the upper bound in (17) or (19); conversely, if we do not want the network to reach a global consensus, we need to take $K$ smaller than the lower bound in (17) (or (19)). In the simple case of unbounded coupling function, i.e. $f_{max} = \infty$, Corollary 1 proves that the critical coupling value is $K = 0$, since $K_L = K_U = 0$.

**Remark 2.** The nonlinear coupling model (1) includes, as a particular case, the linear coupling scheme, corresponding to the choice $f(x) = x$, as this function respects condition **a2**. From (19), we realize that if the function $f(x)$ is linear (and then unbounded), the critical value of $K$ is zero. This means that a linearly coupled system always converges to a synchronized state, for any (positive) value of $K$ (Corollary 1). We may then ask ourselves whether there is any advantage in using a nonlinear as opposed to a linear, system. Indeed, the possibility to switch the system behavior from a system always converging to a global consensus to a system that cannot reach a global consensus, by acting on a single parameter, $K$, is a potential advantage that can be usefully exploited, as will be shown in the next section, to perform some kind of spatial clustering. This variety of behaviors is in fact a unique capability offered by nonlinear, as opposed to linear, systems.

---

[5]The maximum of $f(\cdot)$ is defined on the extended real numbers, i.e., on $\overline{\mathbb{R}} = \mathbb{R} \cup \{-\infty, +\infty\}$.

[6]A function $f : \mathbb{R} \mapsto \mathbb{R}$ is said to be asymptotically convex or concave, if $\exists \overline{x} \in \mathbb{R} : \text{sign}(f''(x)) = \text{sign}(f''(\overline{x})), \forall x \geq \overline{x}$, where $\text{sign}(x)$ denotes the sign of $x$, and $f''(x)$ is the second derivative of $f(x)$ with respect to $x$. Observe that the above condition just avoids that $f(\cdot)$ could change its concavity infinitely often. Thus, it does not represent a strong restriction in the choice of the function $f(\cdot)$.

**Remark 3.** As a by-product of the proof of Theorem 1, in the particular case of $c = 1$, and under the conditions of Theorem 1, the dynamical system (3) approaches the synchronized state with a rate that is locally proportional to $K\lambda_2(\mathbf{L_A})$. The convergence rate is indeed a crucial parameter. In fact, from the point of view of the energy required to reach a common decision, the proposed system has several advantages with respect to centralized systems, as it is totally decentralized, but it has to pay these advantages with the energy wasted in the iterations necessary to reach the estimate. Clearly, the higher is the convergence rate, the lower is this waste of energy. The previous considerations suggest that, to increase the rate, we can increase $K$ or change the network topology in order to increase $\lambda_2(\mathbf{L_A})$, by increasing the network degree, for example.

**Remark 4.** The wireless channel is typically affected by fading. Hence, it is important to analyze the proposed scheme when the coefficients $a_{ij}$ are random variables. From the conditions required from Theorem 1, we realize that, provided that the nonzero fading coefficients are positive (this requirement has an impact on the kind of detector to be used) and the network is connected, the system maintains its capability to achieve a global consensus that is not affected by the values of $a_{ij}$. The only impact of the randomness of the coefficients $a_{ij}$ is on the convergence rate, as $\lambda_2(\mathbf{L_A})$ depends on them.

The properties established by Theorem 1, for scalar observation systems, can be generalized to the vector system (5), as follows.

**Theorem 2** *Given system (4), assume that conditions of Theorem 1 are satisfied. Then, there exist two unique critical values of $K$, denoted by $K_L$ and $K_U$, with $0 \leq K_L \leq K_U$, such that the synchronized state exists for all $K > K_U$, and it does not for all $K < K_L$. Furthermore, if it exists, the synchronized state is also globally asymptotically stable. Upper and lower bounds of $K_L$ and $K_U$ are*

$$K_L \geq \frac{\max_i \|\Delta\bar{\boldsymbol{\omega}}_i\|_\infty}{f_{\max} d_{\max}}, \quad \text{and} \quad K_U \leq \frac{\max_i \|\Delta\bar{\boldsymbol{\omega}}_i\|_2}{g\lambda_2(\mathbf{L_A})}, \tag{20}$$

*where $g$ is defined in (18) and $\Delta\bar{\boldsymbol{\omega}} = [\Delta\bar{\boldsymbol{\omega}}_1^T, \ldots, \Delta\bar{\boldsymbol{\omega}}_L^T] \triangleq \mathbf{PD_Q}(\boldsymbol{\omega} - \mathbf{1}_N \otimes \bar{\boldsymbol{\omega}}_L^\star)$, with $\bar{\boldsymbol{\omega}}_L^\star$ given in (11).*

**Proof.** See Appendix B. ∎

**Corollary 2** *Assume that, $f(\cdot)$, in addition to $\mathbf{a2}$, is asymptotically convex or concave. Then:*

1. *If $f(\cdot)$ is unbounded, $K_L = K_U = 0$;*

2. *If $f(\cdot)$ is bounded, upper and lower bounds of $K_L$ and $K_U$ are*

$$K_L \geq \frac{\max_i \|\Delta\bar{\boldsymbol{\omega}}_i\|_\infty}{f_{\max} d_{\max}}, \quad \text{and} \quad K_U \leq \frac{\max_i \|\Delta\bar{\boldsymbol{\omega}}_i\|_2}{f_{\max}\lambda_2(\mathbf{L_A})}. \tag{21}$$

## 5 Performance

In this section we illustrate some properties of the self-synchronizing network proposed before.

## 5.1 ML estimation in the presence of noise

Different sources of noise affect the system: the observation noise, represented by the vector of random variables $\boldsymbol{w}_i$ in (14), and the system or coupling noise, represented by the stochastic process $v_i(t)$ in (1) (or (4)). These sources of noise affect system performance in a different way, as we show next.

### 5.1.1 Observation Noise

In Section 3, we showed how to choose the network parameters to guarantee the convergence of each node to the globally optimum ML estimate. However, in the case of unbounded noise, the convergence cannot be guaranteed with probability one. In fact, once we have chosen a coefficient $K$, there is always a non null probability that $K_L > K$, event that prevents the possibility of global synchronization. For any chosen $K$, using the upper bound in (19), the probability of the non-synchronization event can be upper bounded as follows:

$$P_{ns} := \mathcal{P}\{K_L > K\} \leq \mathcal{P}\{K_U > K\} \leq \mathcal{P}\left\{\|\boldsymbol{D}_c\Delta\boldsymbol{\omega}\|_2 > \frac{K}{2}f_{max}\lambda_2(\boldsymbol{L}_a)\right\}. \tag{22}$$

In the case of Gaussian observation noise, $\Delta\boldsymbol{\omega}$ is a vector of i.i.d. zero mean Gaussian random variables and thus $\|\boldsymbol{D}_c\Delta\boldsymbol{\omega}\|_2$ is a $\chi$ random variable, with $N$ degrees of freedom. Hence, denoting with $D_Z(z)$ the cumulative distribution function of $Z := \|\boldsymbol{D}_c\Delta\boldsymbol{\omega}\|_2$, the probability that the network does not synchronize, for a given choice of $K$, is upper bounded as

$$P_{ns} \leq 1 - D_Z\left(\frac{K}{2}\ f_{max}\lambda_2(\boldsymbol{L}_A)\right).$$

This probability can be made arbitrarily small by using high values of $K$.

As an example of vector ML estimation, in Fig. 2a), we show the derivatives $\dot{\theta}_i(t)$ as a function of time, for a network composed of $N = 16$ sensors. The nonlinear coupling function $f(x)$ in this case is bounded and equal to $f(x) = \tanh(x)$. Each node has degree 4. The observation model is the linear vector model (14), with $L = 3$, $M = 6$; the matrices $\boldsymbol{A}_i$ are composed of i.i.d. Gaussian random variables of zero mean and unit variance. The common unknown vector is $\boldsymbol{\xi} = [1, 2, 3]^T$. The observation noise is white Gaussian with zero mean and unit variance. The dashed line and the circles represent the global ML estimates achievable with an ideal control node that receives the observations and all mixing matrices $\boldsymbol{A}_i$ with no errors. We can see that, in spite of the low SNR, all the nodes converge to the ML estimate, as predicted. To get a global performance assessment, not conditioned to the specific realizations of the matrices $\boldsymbol{A}_i$, in Fig. 3, we report the variances obtained in the same setting as in Fig.2, averaging over 100 independent realizations of the mixing matrices $\boldsymbol{A}_i$, the sensor initializations $\boldsymbol{\theta}(0)$ and the noise values, as a function of the number of sensors $N$. The network is a regular network, where all nodes have degree 4, for all values of $N$. Fig.3a) refers to the choice $f(x) = \tanh(x)$, whereas Fig.3b) refers to the choice $f(x) = \sin(x)$. The three sets of marks, in each curve, represent the variances obtained with the decentralized ML estimator, whereas the solid lines refer to the centralized ML estimator. The convergence time is fixed to one second and it is the same for all $N$. It is interesting to observe that: 1) even though $\sin(x)$ is not monotonic, and thus it does not satisfy assumption **a2** of Theorem 1, the corresponding system behaves as the system with the monotonically increasing function $\tanh(x)$; 2) the decentralized method has practically the same performance as the centralized one; 3) even though the coupling is only local and it does not change with $N$, the variance decays as $1/N$, as the optimal ML estimator - this confirms the scalability of the proposed approach.

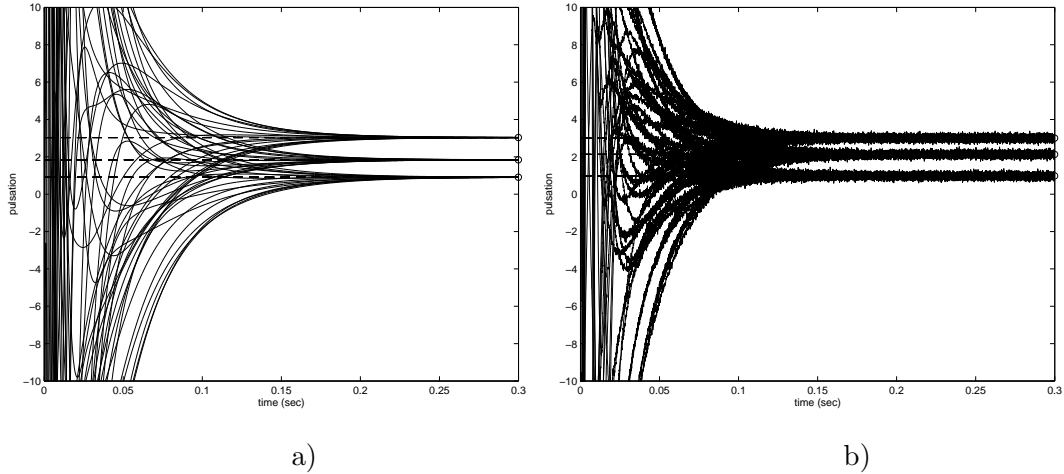

a)                                       b)

Figure 2: Evolution of the pulsation $\dot{\theta}_i(t)$ as a function of time (solid line); optimal centralized ML estimate (dashed lines plus circles): a) observation noise only; b) observation plus coupling noise.

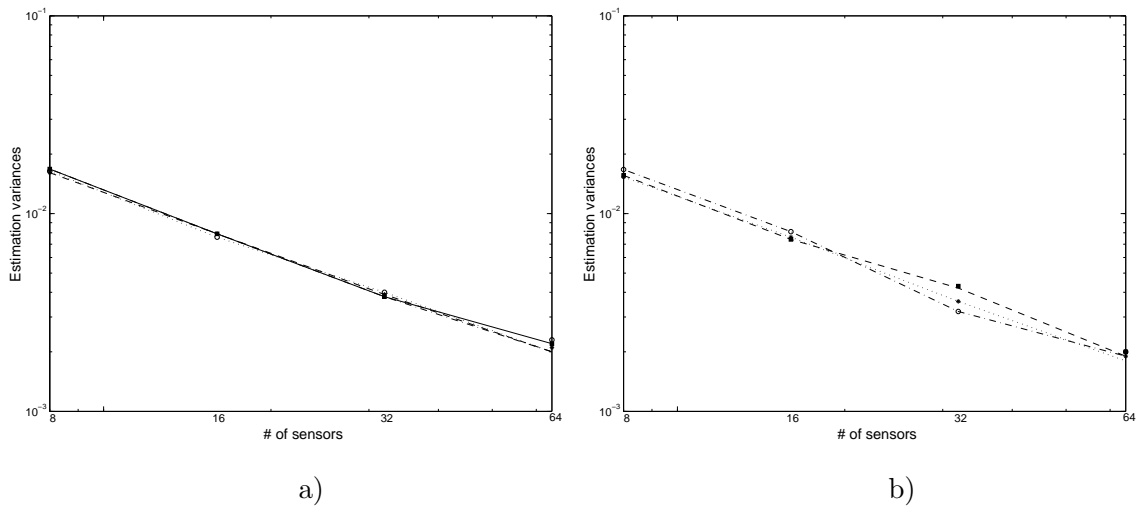

a)                                       b)

Figure 3: Estimation variance as a function of the number of sensors: a) $f(x) = \tanh(x)$; b) $f(x) = \sin(x)$.

### 5.1.2 Coupling Noise

We focus now on the effect of coupling noise on the final estimate. Let us start with the effect of coupling noise on the conventional average consensus algorithms [11, 13, 14, 19, 34]. Without loss of generality, we consider as an example of average consensus algorithm, the discrete-time version of the linearly coupled dynamic system of [11, 29][7]

$$\boldsymbol{x}[n] = \boldsymbol{W}\boldsymbol{x}[n-1] + \boldsymbol{v}[n], n = 1, 2, \ldots, \quad (23)$$

where $\boldsymbol{x}[n] = [x_1[n], \cdots, x_N[n]]^T$, with $x_i[n]$ denoting the scalar state of $i$-th sensor at step $n$, and $\boldsymbol{v}[n]$ is the noise vector at step $n$ and $W_{ij}$ is the weight associated by node $i$ to the signal received from node $j$ ($W_{ij} \neq 0$ if $a_{ij} \neq 0$, i.e., if nodes $i$ and $j$ are connected). If we pre-multiply (23) by $\mathbf{1}^T$ and divide by $N$, we get:

$$\bar{x}[n] = \bar{x}[n-1] + \frac{1}{N}\sum_{i=1}^{N} v_i[n], \quad (24)$$

with $\bar{x}[n] \triangleq (1/N)\sum_{i=1}^{N} x_i[n]$. This shows that the running average $\bar{x}[n]$ undergoes a random walk, thus implying that its variance increases linearly with the time index. This behavior was already observed in [29], whose authors realized that the average consensus achieved through (23) does not converge in any statistical sense (except in the mean). Specifically, in [29] it was shown that, with average consensus algorithms, as given by (23), what converges to a constant value is the variance of the deviations $z_i[n] := x_i[n] - \bar{x}[n]$. To recover from this problem, in [29] it was proposed a very elegant way to minimize the sum of the variances of $z_i[n]$, as a solution of a convex optimization problem, but the running average is still a process with variance increasing with time.

Let us consider now the scalar system (1), but similar results can be obtained for the vectorial case (4). The study of the stability of the dynamical system (1), in the case of nonlinear coupling and in the presence of noise (4) is indeed a difficult problem and it goes beyond the scope of the present paper. Nevertheless, if we limit ourselves to the linear coupling case, i.e. $f(x) = x$, to make a comparison with the average consensus algorithm, as given in (24), we may still derive some basic properties. In fact, in the linear coupling case, exploiting the superposition principle, each dynamical system will converge to a random process having an average value, equal to $\omega^\star$, as given in (9), corresponding to the solution of (1), with $\omega_i \neq 0$ and $v_i(t) = 0$, plus a fluctuating term having zero mean and variance $\sigma_n^2$, corresponding to the solution of (1), with $\omega_i = 0$ and $v_i(t) \neq 0$. In other words, the effect of the coupling noise is to add a noise with constant variance, rather than of increasing variance, on the final estimate. This happens simply because the estimate is associated to the derivative of the state, rather than on the state itself.

As a numerical example, in Fig. 2b), we report a curve referring to the same settings as in Fig. 2a), except that now there is an additive white Gaussian process $v_i(t)$ on each observed component, of zero mean and variance $\sigma_n^2 = 0.1$. We can see that, also in this case, the state derivatives of all the nodes converge to values centered around the globally optimum ML estimates.

---

[7]This model could also be applied, as the discrete-time counterpart of all the works considering average consensus through linear coupling. Furthermore, in the case of nonlinear coupling [13], (23) can be seen as an approximate version, valid when the states are close to each other and the additive noise is small.

## 5.2 Effect of network topology

From (19), it is evident that the synchronization properties depend on the graph topology through the graph algebraic connectivity $\lambda_2(\boldsymbol{L}_A)$. This means that, for a given $K$ and a given number of nodes, different topologies give rise to different behaviors. The easiest case to analyze is that of *regular* graphs, where all the node have the same degree[8]. The algebraic connectivity of an unweighted regular graph of degree $d$, having a ring topology where each node is coupled with only its neighbors, is

$$\lambda_2 = 4 \sum_{i=1}^{d/2} \sin^2(\pi\, i\, /N) \approx \frac{\pi^2 d(d+1)(d+2)}{6N^2}, \qquad (25)$$

where the last approximation is valid for $d \ll N$. In the more general case of non necessarily regular graph, the algebraic connectivity is not known in closed form, but it can be lower bounded as follows [34]

$$\lambda_2 \geq 2\left(1 - \cos\left(\frac{\pi}{N}\right)\right)\delta \approx \frac{\pi^2}{N^2}\delta, \qquad (26)$$

where $\delta$ is the minimum degree and the approximation in (26) is valid for $N \gg 1$. Hence, in both cases, for a given $d$ or $\delta$, if the network size $N$ increases, $\lambda_2$ decreases as $1/N^2$. From (19), this means that to guarantee the self-synchronization, $K$ must increase as $N^2$.

Conversely, *small world* or *scale-free* random graphs exhibit a different behavior. Small worlds graphs exhibit, in fact, for any given degree, larger algebraic connectivity than regular graphs. As far as scale-free graphs are concerned, they are built starting from an initial number of nodes, say $m_0$, and then adding new nodes according to an iterated procedure of growth and preferential attachment [36]. Denoting with $\bar{\lambda}_2(m_0, N)$ the mean value of the second smallest eigenvalue of the Laplacian of a network composed of $N$ nodes, averaged over the graph realizations, it was shown in [35] that the limit of $\bar{\lambda}_2(m_0, N)$ for $N$ going to infinity is constant. This proves that, if the network is built according to a scale-free topology, our strategy respects the *scalability* and *fault tolerance* properties, since, as soon as we choose $K$ larger than the upper bound in (19), for sufficiently large $N$, we are guaranteed that further addition or removal of a few nodes do not affect the global synchronization, and then estimation, capabilities of the network.

## 5.3 Synchrony vs. desynchrony

In Section 4, we showed that if the coupling is nonlinear and $K$ is smaller than a critical value, the system does not converge. In this section, we show an example of the system behavior when we choose the coupling coefficient $K$ in order to avoid the possibility to achieve global consensus. We considered a network of $40 \times 40$ sensors uniformly spaced over a regular planar grid. The initial measurements of the overall grid are reported in Fig. 4 a). Each sensor is initialized with its (noisy) observation and then it evolves according to (1). Each node is coupled with its neighbors and the maximum node degree is 12. A snapshot of the system state, after 1 second, is reported in Fig. 4 b). Comparing Figs. 4 a) and b), we can see that the network is operating a sort of spatial clustering, even though the nodes are keeping evolving in time. The possibility of segmenting the observed field through a population of coupled dynamical systems is still an open research topic that we are currently investigating.

---

[8] The degree of a node is the number of its neighbors, i.e. the number of nodes linked with that node.

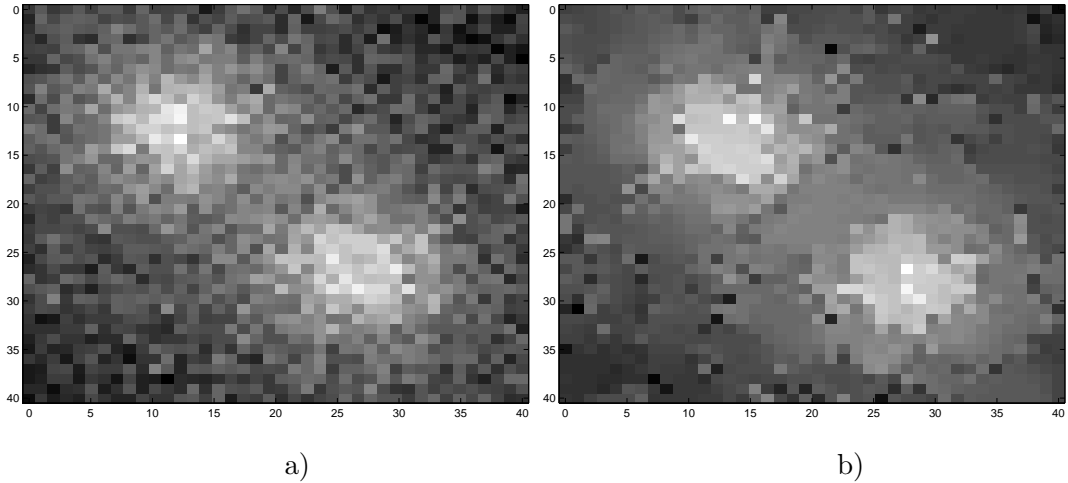

Figure 4: Intensity of the observed field: a) observed field; b) smoothed field.

# 6 Conclusion

In this work we have shown that, if a sensor network observes a common event, a network of nonlinearly coupled first-order dynamical systems can be used to achieve a globally optimum ML estimate, without the need to send any data to any fusion center. We have shown that the conditions guaranteeing the global asymptotic stability of the ML estimate, seen as the self-synchronization state of the whole system, depend on the coupling strength $K$ and on the network topology through the algebraic connectivity $\lambda_2$. With respect to common average consensus techniques, based on the convergence of the state, the approach proposed in this work presents a stronger resilience against additive coupling noise. In general, the major advantages of the proposed strategy, as other average consensus or gossip algorithms, are the simplicity of each node, the scalability of the approach and the absence of congestion problems. These advantages are paid by the fact that the solution is achieved through an iterative procedure, whose convergence rate is proportional to the product $K\lambda_2$. In general, the total energy spent to achieve the final estimate, within a certain accuracy, is proportional to the product between the time necessary to reach the desired estimate (within prescribed error) and the power transmitted by each node. Since the iterative procedure involves only local exchange of information (provided that the overall connectivity is guaranteed), the transmit power of each node may be low. The time necessary to achieve the estimate is inversely proportional to $K\lambda_2(\boldsymbol{L})$. To increase $\lambda_2(\boldsymbol{L})$ it is necessary to increase the node degree, but this entails an increase of the transmit power of each node. Hence, we can foresee a sort of optimal transmit energy, at each node, as a trade-off between the contrasting needs mentioned above. This is indeed an interesting research direction that we are currently investigating. For large scale networks, it is also important to study the effect of propagation delays. In [37] we studied this problem and our results show that, for small delays, the system still converges, but the estimate becomes biased by an amount depending on the delays. Other interesting extensions include the effect of coupling noise for the general nonlinearly coupled system, the effect of having directed graphs, to model a system with different transmit power on each node, the effect of random coupling coefficients, to model channel fading effects, and the impact of time-varying topologies.

# Appendix

## A  Proof of Theorem 1

We first introduce the following intermediate results, that will be used in the proof.

**Lemma 1** *Given an oriented weighted graph $\mathcal{G}$ with $N$ nodes, and positive numbers $\{w_i\}_i$ associated to the edges, let $\mathbf{L_w} \triangleq \mathbf{B}\mathbf{D_w}\mathbf{B}^T$ be the (weighted) Laplacian of $\mathcal{G}$, where $\mathbf{B}$ is the $N \times E$ incidence matrix, $\mathbf{D_w} \triangleq \mathrm{diag}(\mathbf{w})$ is the $E \times E$ diagonal matrix whose diagonal entries are the edge-weights $w_i$. Let $\mathbf{L_w^\sharp}$ denote the generalized inverse of $\mathbf{L_w}$ [38]. If the graph $\mathcal{G}$ is connected, then $\mathbf{L_w^\sharp} : \mathbb{R}_{++}^E \mapsto \mathbb{R}^N$ is a continuous function in $\mathbb{R}_{++}^E$.*

**Proof.** Given $\mathbf{w} \in \mathbb{R}_{++}^E$, consider the eigen-decomposition of $\mathbf{L_w}$

$$\mathbf{L_w} = \mathbf{U_w}\mathbf{\Lambda_w}\mathbf{U_w}^T, \tag{27}$$

with eigenvalues arranged in nondecreasing order. The generalized inverse of $\mathbf{L_w}$ in (27) is given by

$$\mathbf{L_w^\sharp} = \overline{\mathbf{U}}_\mathbf{w}\overline{\mathbf{\Lambda}}_\mathbf{w}^{-1}\overline{\mathbf{U}}_\mathbf{w}^T, \tag{28}$$

where $\overline{\mathbf{\Lambda}}_\mathbf{w}$ is the $r \times r$ diagonal matrix, containing the *last $r$* positive eigenvalues of $\mathbf{L_w}$, and $\overline{\mathbf{U}}_\mathbf{w}$ the $N \times r$ matrix of the corresponding eigenvectors. Since the graph $\mathcal{G}$ is assumed to be connected, we have $r = N - 1$, i.e. [31, Lemma 13.1.1]

$$\mathrm{rank}(\mathbf{\Lambda_w}) = N - 1, \quad \forall \mathbf{w} \in \mathbb{R}_{++}^n. \tag{29}$$

The generalized inverse $\mathbf{L_w^\sharp}$ in (28) is continuous in $\mathbb{R}_{++}^E$ if it is continuous at $\mathbf{w}$, for any fixed $\mathbf{w} \in \mathbb{R}_{++}^E$. Given $\mathbf{w} \in \mathbb{R}_{++}^E$, $\mathbf{L_w^\sharp}$ is continuous at $\mathbf{w}$ if, for any sequence $\{\mathbf{w}_j\}_{j \in \mathbb{N}}$ of positive vectors $\mathbf{w}_j$ that converges to $\mathbf{w}$, the corresponding sequence of generalized inverses $\{\mathbf{L}_{\mathbf{w}_j}^{\sharp}\}_{j\in\mathbb{N}}$ converges to $\mathbf{L}_{\mathbf{w}}^{\sharp}$[9]. Stated in mathematical terms, we need to prove that

$$\forall \{\mathbf{w}_j\}_{j\in\mathbb{N}} \to \mathbf{w} \quad \Rightarrow \quad \{\mathbf{L}_{\mathbf{w}_j}^{\sharp}\}_{j\in\mathbb{N}} \to \mathbf{L}_{\mathbf{w}}^{\sharp}, \tag{30}$$

where $\mathbf{L}_{\mathbf{w}_j}^{\sharp}$ denotes the generalized inverse of the weighted Laplacian $\mathbf{L}_{\mathbf{w}_j}$ associated to the weights-vector $\mathbf{w}_j$. Since $\mathbf{L}_{\mathbf{w}}$ is a continuous function of $\mathbf{w}$, (30) is equivalent to the following

$$\forall \{\mathbf{L}_{\mathbf{w}_j}\}_{j\in\mathbb{N}} \to \mathbf{L}_{\mathbf{w}} \quad \Rightarrow \quad \{\mathbf{L}_{\mathbf{w}_j}^{\sharp}\}_{j\in\mathbb{N}} \to \mathbf{L}_{\mathbf{w}}^{\sharp}. \tag{31}$$

We prove now that (31) is satisfied, provided that (29) holds true (i.e., the graph $\mathscr{G}$ is connected). To this end, we use the following necessary and sufficient condition for the continuity of the generalized (Drazin) inverse [38, Definition 7.2.3] [10].

**Theorem 3** ([38, Theorem 10.7.1][39, Theorem 2]) *Let* $\{\mathbf{L}_{\mathbf{w}_j}\}_{j\in\mathbb{N}} \to \mathbf{L}_{\mathbf{w}}$, *with* $\mathbf{L}_{\mathbf{w}}$ *and* $\mathbf{L}_{\mathbf{w}_j}$ *given by (27). Then,* $\{\mathbf{L}_{\mathbf{w}_j}^{\sharp}\}_{j\in\mathbb{N}} \to \mathbf{L}_{\mathbf{w}}^{\sharp}$ *if and only if*

$$\exists j_0 > 0 \; : \; \mathrm{rank}(\boldsymbol{\Lambda}_{\mathbf{w}_j}) = \mathrm{rank}(\boldsymbol{\Lambda}_{\mathbf{w}}), \quad \forall j \geq j_0. \tag{32}$$

If the graph is connected, from (29) it follows that $\mathrm{rank}(\boldsymbol{\Lambda}_{\mathbf{w}_j}) = \mathrm{rank}(\boldsymbol{\Lambda}_{\mathbf{w}})$, for all $\mathbf{w}, \mathbf{w}_j \in \mathbb{R}_{++}^E$. Thus, given $\mathbf{w} \in \mathbb{R}_{++}^E$, for any convergent sequence $\{\mathbf{L}_{\mathbf{w}_j}\}_{j\in\mathbb{N}} \to \mathbf{L}_{\mathbf{w}}$ in (31), condition (32) is satisfied, and hence $\{\mathbf{L}_{\mathbf{w}_j}^{\sharp}\}_{j\in\mathbb{N}} \to \mathbf{L}_{\mathbf{w}}^{\sharp}$. This proves the continuity of $\mathbf{L}_{\mathbf{w}}^{\sharp}$ at $\mathbf{w} \in \mathbb{R}_{++}^E$. Since condition (31) is satisfied for any $\mathbf{w} \in \mathbb{R}_{++}^E$, $\mathbf{L}_{\mathbf{w}}^{\sharp}$ is continuous in all $\mathbb{R}_{++}^E$. ∎

**Lemma 2** ([40, Theorem 4.14]) *Let $\mathcal{C}$ be a closed, convex subset of a normed linear space. Then, every compact[11], continuous map $F : \mathcal{C} \mapsto \mathcal{C}$ admits at least one fixed point.*

The proof of Theorem 1 is organized as follows. We introduce, first, a proper transformation of the original system (3), so that the existence and the *global asymptotic* stability (according to Definition 1) of the synchronized state (9) can be recasted in the classical study of existence and the *asymptotic* stability of the equilibria of the transformed system (see, e.g., [41, 42]). Then, using standard fixed point arguments, we prove that, under **a1-a3**, an equilibrium for the transformed system exists, provided that $K > K_U$, and cannot exists if $K < K_L$. Finally, we show, introducing a valid Lyapunov function, that, if an equilibrium exists, it is also asymptotically stable.

---

[9] A sequence $\{\mathbf{A}_j\}_{j\in\mathbb{N}}$ of matrices in $\mathbb{R}^{n\times n}$ is said to converge to $\mathbf{A} \in \mathbb{R}^{n\times n}$, if for every real number $\epsilon > 0$, there exists an index $j_0(\epsilon)$ such that if $j \geq j_0(\epsilon)$, then $\|\mathbf{A}_j - \mathbf{A}\| < \epsilon$, where $\|\cdot\|$ is some matrix norm on $\mathbb{R}^{n\times n}$. We denote a convergent sequence $\{\mathbf{A}_j\}_{j\in\mathbb{N}}$ to $\mathbf{A}$ as $\{\mathbf{A}_j\}_{j\in\mathbb{N}} \to \mathbf{A}$. It is worth observing that $\{\mathbf{A}_j\}_{j\in\mathbb{N}} \to \mathbf{A}$ if and only if the entries of $\mathbf{A}_j$ converge to the corresponding entries of $\mathbf{A}$.

[10] For the sake of simplicity, we adapt [38, Theorem 10.7.1] to our notation. Observe that the Drazin inverse, as defined in [38, Definition 7.2.3], corresponds to the generalized inverse given in (28).

[11] The map $f : \mathcal{C} \mapsto \mathcal{C}$ is called compact if $f(\mathcal{C})$ is contained in a compact subset of $\mathcal{C}$.

## A.1 Existence of the synchronized state

Let us assume that conditions **a1-a3** are satisfied and consider the following change of variables

$$\Psi_i(t) = \theta_i(t) - \omega^\star t, \quad i = 1, 2, \ldots, N, \tag{33}$$

with $\omega^\star$ defined in (9). The original system (3) can be equivalently rewritten as

$$\dot{\boldsymbol{\Psi}}(t) = \Delta\boldsymbol{\omega} - K\,\mathbf{D_c}^{-1}\mathbf{B}\mathbf{D_A} f\left[\mathbf{B}^T\boldsymbol{\Psi}(t)\right] \tag{34}$$

$$= \Delta\boldsymbol{\omega} - K\mathbf{D_c}^{-1}\mathbf{B}\mathbf{D_A}\mathbf{D_{\Psi}}\mathbf{B}^T\boldsymbol{\Psi}(t) \tag{35}$$

$$\triangleq \Delta\boldsymbol{\omega} - K\mathbf{D_c}^{-1}\mathbf{L_{A,\Psi}}\boldsymbol{\Psi}(t), \tag{36}$$

where $\boldsymbol{\Psi}(t) \triangleq [\Psi_1(t), \ldots, \Psi_N(t)]^T$, with $\boldsymbol{\Psi}(0) = \boldsymbol{\theta}(0)$, $\Delta\boldsymbol{\omega} = \boldsymbol{\omega} - \omega^\star \mathbf{1}$, and $\mathbf{L_{A,\Psi}}$ is the weighted Laplacian of the graph, with diagonal weights-matrix $\mathbf{D_A}\mathbf{D_{\Psi}}$ (that depends on $\boldsymbol{\Psi}$), and $[\mathbf{D_{\Psi}}]_{ii}$ is given by

$$[\mathbf{D_{\Psi}}]_{ii} \triangleq \frac{f\left([\mathbf{B}^T\boldsymbol{\Psi}]_i\right)}{[\mathbf{B}^T\boldsymbol{\Psi}]_i} > 0, \quad \forall \boldsymbol{\Psi} \in \mathbb{R}^N, \quad i = 1, \ldots, E, \tag{37}$$

where the positivity of $[\mathbf{D_{\Psi}}]_{ii} > 0$ comes from **a1**[12].

According to Definition 1, the synchronized state of (3) exists if and only if the dynamical system in (36) admits an equilibrium, or equivalently, if there exists a solution for the following system of nonlinear equations

$$\mathbf{L_{A,\Psi}}\boldsymbol{\Psi} = \frac{1}{K}\mathbf{D_c}\Delta\boldsymbol{\omega}. \tag{38}$$

Given (38), we prove that there exist two critical non-negative[13] values of $K$, denoted by $K_L$ and $K_U$, such that for *all* $K > K_U$ the system (38) is feasible and for *all* $K < K_L$ the solution of (38) disappears. To this end, it is sufficient to provide a lower bound $K_L^{(\text{low})}$ of $K_L$, and an upper bound $K_U^{(\text{up})}$ of $K_U$, such that for *all* $K \geq K_U^{(\text{up})}$ the system (38) admits a solution, and for all $K < K_L^{(\text{low})}$ does not. Given these $K_L^{(\text{low})}$ and $K_U^{(\text{up})}$, there must exist a unique $K_L$ and $K_U$ as defined above such that $K_L^{(\text{low})} \leq K_L \leq K_U \leq K_U^{(\text{up})}$.

It is worth observing that, in the case of the linear function $f(x) = x$, we have $[\mathbf{D_{\Psi}}]_{ii} = 1$ and $\mathbf{L_{A,\Psi}} = \mathbf{L_A}$, $\forall \boldsymbol{\Psi} \in \mathbb{R}^N$. Hence, the system (38) becomes linear and, since $\mathbf{1}_N^T \mathbf{D_c}\Delta\boldsymbol{\omega} = 0$ and $\mathbf{L_{A,\Psi}}\mathbf{1}_N = \mathbf{0}_N$, it admits, for any $K \neq 0$, $\infty^1$ solutions, given by $\boldsymbol{\Psi}^\star = \mathbf{L_A}^\sharp \mathbf{D_c}\Delta\boldsymbol{\omega}/K + \text{span}\{\mathbf{1}_N\}$, where $\mathbf{L_A}^\sharp$ is the generalized inverse of the weighted Laplacian $\mathbf{L_A}$ [38]. Thus, in the case of $f(x) = x$, we have $K_U = 0$ (by definition) and $K_L = 0$ (since for $K < 0$ the equilibrium points of (36) are not stable, as shown in Appendix A.2).

---
[12] Note, from item 3, right after (1), that $\lim_{x \to 0} f(x)/x = \dot{f}(0) = 1$.

[13] We focus only on nonnegative $K$, since the potential solutions of the system (38) corresponding to negative values of $K$ are not stable for (3), as we will show in Appendix A.2.

Conversely, in the case of nonlinear bounded functions $f(\cdot)$, i.e. $\lim_{x \to +\infty} f(x) = f_{\max} < +\infty$, $K_L$ is lower bounded by a positive quantity. This lower bound corresponds to the value of $K$ below which the system (38) is surely infeasible, i.e. (see (34))

$$K \left\| \mathbf{B} \mathbf{D_A} f \left[ \mathbf{B}^T \boldsymbol{\Psi} \right] \right\|_\infty < \left\| \mathbf{D_c} \Delta \boldsymbol{\omega} \right\|_\infty, \tag{39}$$

where $\|\cdot\|_\infty$ denotes the infinity norm of a vector. A more stringent condition than (39) is

$$K d_{\max} f_{\max} < \left\| \mathbf{D_c} \Delta \boldsymbol{\omega} \right\|_\infty, \tag{40}$$

where we used the inequality[14] $\|\mathbf{B}\mathbf{D_A}\|_\infty \leq d_{\max}$, with $d_{\max} \triangleq \max_i \sum_{j=1}^{E} [\mathbf{B}\mathbf{D_A}]_{ij}$. From (40) it follows that, if the system (38) admits a solution, then it must be

$$K \geq K_L^{(\text{low})} \triangleq \frac{\|\mathbf{D_c} \Delta \boldsymbol{\omega}\|_\infty}{d_{\max} f_{\max}}, \tag{41}$$

which provides the lower bound in (17). Observe that, for unbounded nonlinear functions (i.e. $f_{\max} = +\infty$), the lower bound (41) disappears.

We consider now a generic (bounded or unbounded) *nonlinear* function $f(\cdot)$, and provide sufficient conditions on $K$ for the system (38) to be feasible. A solution for (38) exists, if and only if the following mapping admits at least one fixed point in $\mathbb{R}^N$

$$\boldsymbol{\Psi} = \frac{\mathbf{L}^\sharp_{\mathbf{A},\boldsymbol{\Psi}}}{K} \mathbf{D_c} \Delta \boldsymbol{\omega}, \tag{42}$$

where $\mathbf{L}^\sharp_{\mathbf{A},\boldsymbol{\Psi}}$ is the generalized inverse of the weighted Laplacian $\mathbf{L}_{\mathbf{A},\boldsymbol{\Psi}}$, given by

$$\mathbf{L}^\sharp_{\mathbf{A},\boldsymbol{\Psi}} = \overline{\mathbf{U}}_{\mathbf{A},\boldsymbol{\Psi}} \overline{\boldsymbol{\Lambda}}^{-1}_{\mathbf{A},\boldsymbol{\Psi}} \overline{\mathbf{U}}^T_{\mathbf{A},\boldsymbol{\Psi}}, \tag{43}$$

with $\overline{\boldsymbol{\Lambda}}_{\mathbf{A},\boldsymbol{\Psi}}$ the $(N-1) \times (N-1)$ diagonal matrix, containing the *last* $N-1$ positive eigenvalues of $\mathbf{L}_{\mathbf{A},\boldsymbol{\Psi}}$ (assumed to be arranged in nondecreasing order), and $\overline{\mathbf{U}}_{\mathbf{A},\boldsymbol{\Psi}}$ the $N \times (N-1)$ matrix of the corresponding eigenvectors. To prove the existence of at least one fixed-point for (42) it is sufficient to show that (42) admits a solution in some compact, convex set of $\mathbb{R}^N$. We choose this set, without loss of generality (w.l.o.g.) as $\mathcal{B}_a \triangleq \{\boldsymbol{\Psi} \in \mathbb{R}^N : \|\boldsymbol{\Psi}\|_2 \leq a\}$, where $a$ is any positive number $\mathbb{R}_{++}$. Observe that $\mathcal{B}_a$ is compact and convex for all $a \in \mathbb{R}_{++}$, and that, by Lemma 1, the mapping $\mathbf{L}^\sharp_{\mathbf{A},\boldsymbol{\Psi}} \mathbf{D_c} \Delta \boldsymbol{\omega}/K$ in (42) is continuous on $\mathbb{R}^N$ (because of the positivity of $[\mathbf{D}_{\boldsymbol{\Psi}}]_{ii} > 0, \forall \boldsymbol{\Psi} \in \mathbb{R}^N$). Hence, according to Lemma 2, a fixed-point for (42) exists in $\mathcal{B}_a$, if $\mathbf{L}^\sharp_{\mathbf{A},\boldsymbol{\Psi}} \mathbf{D_c} \Delta \boldsymbol{\omega}/K$ is a compact map on $\mathcal{B}_a$, for some given $a \in \mathbb{R}_{++}$. This is guaranteed if, for any given $a \in \mathbb{R}_{++}$, $K$ is chosen so that $\|\mathbf{L}^\sharp_{\mathbf{A},\boldsymbol{\Psi}} \mathbf{D_c} \Delta \boldsymbol{\omega}\|_2 \leq K a$, which corresponds to

$$K \geq \frac{\|\mathbf{L}^\sharp_{\mathbf{A},\boldsymbol{\Psi}}\|_2 \|\mathbf{D_c} \Delta \boldsymbol{\omega}\|_2}{a} = \frac{\|\mathbf{D_c} \Delta \boldsymbol{\omega}\|_2}{a \, \lambda_2 (\mathbf{L}_{\mathbf{A},\boldsymbol{\Psi}})}, \qquad \forall \boldsymbol{\Psi} \in \mathcal{B}_a, \text{ and } a \in \mathbb{R}_{++}, \tag{44}$$

---

[14]The matrix norm induced by the vector infinity norm is the maximum among the absolute values of the row sums [38, Proposition 10.2.2].

where $\|\mathbf{L}_{\mathbf{A},\boldsymbol{\Psi}}^{\sharp}\|_2$ is the spectral norm of $\mathbf{L}_{\mathbf{A},\boldsymbol{\Psi}}^{\sharp}$[15] and $\lambda_2(\mathbf{L}_{\mathbf{A},\boldsymbol{\Psi}})$ is the second smallest eigenvalue of $\mathbf{L}_{\mathbf{A},\boldsymbol{\Psi}}$. To remove the dependence of $\lambda_2(\mathbf{L}_{\mathbf{A},\boldsymbol{\Psi}})$ on $\boldsymbol{\Psi}$, we consider the more stringent (sufficient) condition

$$\frac{\|\mathbf{D_c}\Delta\boldsymbol{\omega}\|_2}{a\,\lambda_2(\mathbf{L}_{\mathbf{A},\boldsymbol{\Psi}})} \leq \frac{\|\mathbf{D_c}\Delta\boldsymbol{\omega}\|_2}{a\min_{\boldsymbol{\Psi}\in\mathcal{C}_a}\lambda_2(\mathbf{L}_{\mathbf{A},\boldsymbol{\Psi}})} \leq K, \qquad a\in\mathbb{R}_{++}, \tag{45}$$

where we used the inequality $\min_{\boldsymbol{\Psi}\in\mathcal{C}_a}\lambda_2(\mathbf{L}_{\mathbf{A},\boldsymbol{\Psi}}) \leq \min_{\boldsymbol{\Psi}\in\mathcal{B}_a}\lambda_2(\mathbf{L}_{\mathbf{A},\boldsymbol{\Psi}})$, with $\mathcal{C}_a \triangleq [-a,a]^N \supseteq \mathcal{B}_a$ [43, Theorem 1.16]. In (45), the minimum of $\lambda_2(\mathbf{L}_{\mathbf{A},\boldsymbol{\Psi}})$ on $\mathcal{C}_a$ can be lower bounded as follows. For all $\overline{\boldsymbol{\Psi}} \geq \widetilde{\boldsymbol{\Psi}} \geq \mathbf{0}$, since $\mathbf{L}_{\mathbf{A},\overline{\boldsymbol{\Psi}}} - \mathbf{L}_{\mathbf{A},\widetilde{\boldsymbol{\Psi}}} \succeq \mathbf{0}$, we have [44, Problem 4.3.14]

$$\lambda_i(\mathbf{L}_{\mathbf{A},\overline{\boldsymbol{\Psi}}}) \geq \lambda_i(\mathbf{L}_{\mathbf{A},\widetilde{\boldsymbol{\Psi}}}), \quad \forall\overline{\boldsymbol{\Psi}} \geq \widetilde{\boldsymbol{\Psi}} \geq \mathbf{0}_N, \quad i=1,\ldots,N, \tag{46}$$

where $\lambda_i(\mathbf{L}_{\mathbf{A},\overline{\boldsymbol{\Psi}}})$ and $\lambda_i(\mathbf{L}_{\mathbf{A},\widetilde{\boldsymbol{\Psi}}})$ denote the $i$-th eigenvalue of $\mathbf{L}_{\mathbf{A},\overline{\boldsymbol{\Psi}}}$ and $\mathbf{L}_{\mathbf{A},\widetilde{\boldsymbol{\Psi}}}$, respectively, arranged according to the same order. From (46), it follows that the lower bound of the minimum of $\lambda_2(\mathbf{L}_{\mathbf{A},\boldsymbol{\Psi}})$ on $\mathcal{C}_a$ occurs for the minimum of the weights $[\mathbf{D}_{\boldsymbol{\Psi}}]_{ii}$, defined in (37). Since $[\mathbf{B}^T\boldsymbol{\Psi}]_i \in [-2a,2a]$ for any $\boldsymbol{\Psi}\in\mathcal{C}_a$, we have

$$\lambda_2(\mathbf{L}_{\mathbf{A},\boldsymbol{\Psi}}) \geq \lambda_2(\mathbf{L}_\mathbf{A})\min_{x\in[-2a,2a]}\frac{f(x)}{x} = \lambda_2(\mathbf{L}_\mathbf{A})\min_{x\in[0,2a]}\frac{f(x)}{x}, \qquad \forall\boldsymbol{\Psi}\in\mathcal{C}_a, \tag{47}$$

where the last equality follows from the fact that, because of **a1**, $f(x)/x$ is even. Using (45) and (47), we obtain the following bound for $K$:

$$K \geq K(a) \triangleq \frac{1}{a\min_{x\in[0,2a]}\frac{f(x)}{x}}\frac{\|\mathbf{D_c}\Delta\boldsymbol{\omega}\|_2}{\lambda_2(\mathbf{L}_\mathbf{A})}, \qquad a\in\mathbb{R}_{++}. \tag{48}$$

For *any* given $a\in\mathbb{R}_{++}$, $K(a)$ defined in (48) represents the smallest value of $K$, for which a solution of (42) is guaranteed to exists in $\mathcal{B}_a$. Increasing $a\in\mathbb{R}_{++}$, we enlarge the region $\mathcal{B}_a$ where the solution may fall. Since *any* $K(a)$ in (48), with $a\in\mathbb{R}_{++}$, is a valid upper bound for $K_U$, the lowest upper bound of $K_U$ is obtained taking the greatest lower bound (i.e. the infimum) of the set

$$\mathcal{K} \triangleq \{K(a) : a\in\mathbb{R}_{++}\}, \tag{49}$$

with $K(a)$ given by (48). In fact, by the approximation property of the infimum [43, Theorem 1.14], for every $K > \inf\mathcal{K}$, there always exists some $K(a)\in\mathcal{K}$ such that $K \geq K(a) > \inf\mathcal{K}$, which, by (48), is sufficient to guarantee a solution of (42). Since $\mathcal{K}$ in (49) is bounded from below and non-empty, $\inf\mathcal{K}$ always exists [43, Axiom 10] and is given by

$$\inf\mathcal{K} = \frac{\|\mathbf{D_c}\Delta\boldsymbol{\omega}\|_2}{g\lambda_2(\mathbf{L}_\mathbf{A})}, \tag{50}$$

where[16]

$$g = \sup\left\{a\min_{x\in[0,2a]}\frac{f(x)}{x},\quad a\in\mathbb{R}_{++}\right\}. \tag{51}$$

---

[15] The matrix norm consistent with the Euclidean vector norm is the spectral norm [38, Proposition 10.2.4], defined as the largest singular value of the matrix.

[16] Note that the sup in (51) is defined on the extended real numbers, i.e. on $\overline{\mathbb{R}} = \mathbb{R}\cup\{-\infty,+\infty\}$.

Since $\min_{x \in [0,2a]} f(x)/x \leq f(2a)/(2a)$, $\forall a \in \mathbb{R}_{++}$, expression (51) can be upper bounded by [43, Theorem 1.16]

$$g = \sup \left\{ a \min_{x \in [0,2a]} \frac{f(x)}{x}, \quad a \in \mathbb{R}_{++} \right\} \leq \sup \left\{ a \frac{f(2a)}{2a}, \quad a \in \mathbb{R}_{++} \right\} = \frac{f_{\max}}{2}, \qquad (52)$$

which provides the following lower bound for $\inf \mathcal{K}$ in (50):

$$\inf \mathcal{K} \geq \frac{2 \|\mathbf{D_c} \Delta \boldsymbol{\omega}\|_2}{f_{\max} \lambda_2 (\mathbf{L_A})}. \qquad (53)$$

This complete the proof of (17).

Observe that the lower bound in (53) may be reached or not, depending on the particular function $f(x)$. Without additional properties on $f(\cdot)$, we have no guarantee about the achievability.

We prove now that a sufficient condition for $\inf \mathcal{K}$ to satisfy (53) with the equality is that $f(\cdot)$ be asymptotically convex or concave. Stated in mathematical terms, a function $f(\cdot)$ is asymptotically convex or concave if

**a4**) : $\qquad \exists \, \overline{x} \in \mathbb{R} : \text{sign}\left(f''(x)\right) = \text{sign}\left(f''(\overline{x})\right), \quad \forall x \geq \overline{x},$ (54)

where $\text{sign}(x)$ denotes the sign of $x$, and $f''(x)$ is the second derivative of $f(x)$ with respect to $x$.

Under **a1**-**a4**, the function $f(x)/x$ is continuous on $\mathbb{R}$ and it is quasi-convex or quasi-concave on $[\overline{x}, +\infty)$, where $\overline{x}$ is defined in (54). In fact, because of **a4**, $f(x)$ is convex (or concave) on $[\overline{x}, +\infty)$, and thus the set $\{x \in [\overline{x}, +\infty) : f(x) - \alpha x \leq 0\}$ (or $\{x \in [\overline{x}, +\infty) : f(x) - \alpha x \geq 0\}$ ), with $\alpha \in \mathbb{R}$, is convex, because it is the sublevel set (or the superlevel set) of the convex (or concave) function $f(x) - \alpha x$; which corresponds to the definition of quasi-convexity (or quasi-concavity) of $f(x)/x$ on $[\overline{x}, +\infty)$ [45]. Hence, one of the two following statements must hold for $f(x)/x$ on $[\overline{x}, +\infty)$ [45, Section 3.4.2][17]:

$$\exists \, \widetilde{x} \in [\overline{x}, +\infty) : \forall x \in [\widetilde{x}, +\infty), \, f(x)/x \text{ is nondecreasing}, \qquad (55)$$

if and only if $f(x)/x$ is quasi-convex on $[\overline{x}, +\infty)$, or

$$\exists \, \widetilde{x} \in [\overline{x}, +\infty) : \forall x \in [\widetilde{x}, +\infty), \, f(x)/x \text{ is nonincreasing}, \qquad (56)$$

if and only $f(x)/x$ is quasi-concave on $[\overline{x}, +\infty)$. Using (55) and (56), we obtain the following result for the minimum of $f(x)/x$ on $[0, 2a]$[18]:

- If $f(x)/x$ is quasi-convex on $[\overline{x}, +\infty)$, then (see (55))

$$\exists \, \widetilde{a} \in [\overline{x}, +\infty) : \min_{x \in [0,2a]} \frac{f(x)}{x} = \frac{f(w)}{w}, \, \forall a \in [\widetilde{a}/2, +\infty) \text{ and some } w \in [0, \widetilde{a}]; \qquad (57)$$

---
[17] The class of functions $f(x)/x$ that are nonincreasing (or nondecreasing) on the whole interval $[\overline{x}, +\infty]$ is tacitly treated as special case of (55) and (56), as shown after (57) and (58).

[18] We assumed, w.l.o.g., that $\widetilde{a} \geq 0$, since, if $\widetilde{a} < 0$, conditions (55) and (56) are satisfied for any $x \geq 0$.

- If $f(x)/x$ is quasi-concave, either (57) or the following condition hold true (see (56))

$$\exists \, \widetilde{a} \in [\overline{x}, +\infty) : \min_{x \in [0, 2a]} \frac{f(x)}{x} = \frac{f(2a)}{2a}, \quad \forall a \in [\widetilde{a}/2, +\infty). \tag{58}$$

Observe that, in the case where $f(x)/x$ (satisfying (54)) is nondecreasing (nonincreasing) on the whole interval $[\overline{x}, +\infty)$, condition (57) (condition (58)) still holds true. For a bounded function $f(\cdot)$, $\lim_{x \to +\infty} f(x)/x = 0$, and thus (only) condition (58) is satisfied. In the case of unbounded $f(\cdot)$, instead, either (57) or (58) may be met.

Using (57) and (58), we show that, under **a4**, $\inf \mathcal{K}$ defined in (50) achieves the lower bound in (53). To this end, consider the following subset of $\mathcal{K}$

$$\mathcal{K}_{\widetilde{a}} \triangleq \{K(a) : a \in [\widetilde{a}, +\infty)\} \subseteq \mathcal{K}, \tag{59}$$

where $K(a)$ is given in (48) and $\widetilde{a}$ is implicitly defined in (57), if $f(x)$ is asymptotically convex, or in (58), if $f(x)$ is asymptotically concave.[19] Given (57) and (58), $K(a) \in \mathcal{K}_{\widetilde{a}}$ can be written as

$$K(a) = \begin{cases} \dfrac{\|\mathbf{D_c}\Delta\boldsymbol{\omega}\|_2}{a\dfrac{f(w)}{w}\lambda_2(\mathbf{L_A})}, & \text{if (57) holds,} \\[2mm] \dfrac{2\|\mathbf{D_c}\Delta\boldsymbol{\omega}\|_2}{f(2a)\lambda_2(\mathbf{L_A})}, & \text{if (58) holds,} \end{cases} \quad \forall a \in [\widetilde{a}/2, +\infty). \tag{60}$$

Whether (57) or (58) is satisfied, $K(a)$ in (60) is a continuous and decreasing function on $[\widetilde{a}/2, +\infty)$. It follows that

$$\inf \mathcal{K}_{\widetilde{a}} = \lim_{\substack{a \to +\infty \\ a \geq \widetilde{a}}} K(a) = \begin{cases} 0, & \text{if } f(\cdot) \text{ is unbounded,} \\[2mm] \dfrac{2\|\mathbf{D_c}\Delta\boldsymbol{\omega}\|_2}{f_{\max}\lambda_2(\mathbf{L_A})}, & \text{if } f(\cdot) \text{ is bounded.} \end{cases} \tag{61}$$

Using (53), (59) and (61) we obtain[20]

$$\frac{2\|\mathbf{D_c}\Delta\boldsymbol{\omega}\|_2}{f_{\max}\lambda_2(\mathbf{L_A})} \leq \inf \mathcal{K} \leq \inf \mathcal{K}_{\widetilde{a}} = \frac{2\|\mathbf{D_c}\Delta\boldsymbol{\omega}\|_2}{f_{\max}\lambda_2(\mathbf{L_A})}, \tag{62}$$

where the second inequality in (62) comes out from $\mathcal{K}_{\widetilde{a}} \subseteq \mathcal{K}$. From (62) it follows that $\inf \mathcal{K} = \inf \mathcal{K}_{\widetilde{a}}$, which proves the upper bound in (19).

## A.2 Global stability of the synchronized state

Assume now that, in addition to conditions **a1** and **a3**, $K > K_U \geq 0$, so that system (3) may synchronize. We prove that the synchronized state of the system (3), whose existence is guaranteed by $K > K_U$, is globally asymptotically stable (according to Definition 1).

---

[19] With a slight abuse of notation, we use the same symbol $\widetilde{a}$ for both of the conditions (57) and (58).

[20] We use a unified expression for $\inf \mathcal{K}_{\widetilde{a}}$, for both bounded and unbounded functions $f(\cdot)$, with the convention that, if $f_{\max} = +\infty$, then $\inf \mathcal{K}_{\widetilde{a}} = 0$.

To this end, it is sufficient to consider system (34) and show that the state of (34) converges to an equilibrium, from any set of initial conditions. Left-multiplying (34) by $\mathbf{1}_N^T \mathbf{D_c}$ and using (2), we obtain

$$\mathbf{1}_N^T \mathbf{D_c} \dot{\mathbf{\Psi}}(t) = \mathbf{1}_N^T \mathbf{D_c} \Delta \boldsymbol{\omega} - K \mathbf{1}_N^T \mathbf{B} \mathbf{D_A} f\left(\mathbf{B}^T \mathbf{\Psi}(t)\right) = 0 \quad \Leftrightarrow \quad \mathbf{1}_N^T \mathbf{D_c} \mathbf{\Psi}(t) = \mathbf{1}_N^T \mathbf{D_c} \mathbf{\Psi}(0), \quad \forall t \geq 0. \quad (63)$$

In words, the weighted sum $\mathbf{1}_N^T \mathbf{D_c} \mathbf{\Psi}(t)$ is an invariant for system (34). The invariance of $\mathbf{1}_N^T \mathbf{D_c} \mathbf{\Psi}(t)$ allows the following decomposition of the state vector $\mathbf{\Psi}(t)$ in (34)

$$\mathbf{\Psi}(t) \triangleq \frac{\mathbf{1}_N^T \mathbf{D_c} \mathbf{\Psi}(0)}{\mathbf{1}_N^T \mathbf{c}} \mathbf{1}_N + \mathbf{\Delta}(t), \quad (64)$$

with

$$\mathbf{1}_N^T \mathbf{D_c} \mathbf{\Delta}(t) = 0. \quad (65)$$

Thus, we can study the evolution of system (36) by studying the dynamics of the following system

$$\dot{\mathbf{\Delta}}(t) = \Delta \boldsymbol{\omega} - K \mathbf{D_c}^{-1} \mathbf{B} \mathbf{D_A} f\left(\mathbf{B}^T \mathbf{\Delta}(t)\right), \quad (66)$$

with $\mathbf{\Delta}(t)$ satisfying the constraint (65).

We focus now on the equilibria of (65) and show that system (66) admits a unique equilibrium and that such an equilibrium is globally asymptotically stable; this proves also the globally asymptotic stability of the synchronized state of (3).

Since $\mathbf{c}^T \mathbf{1}_N \neq 0$, the condition $K > K_U$ guarantees the existence of an equilibrium for system (66) (c.f. Appendix A.1). Moreover, because of (65), all the equilibria of (66) are isolated[21]. In fact, the Jacobian of $\mathbf{B} \mathbf{D_A} f\left(\mathbf{B}^T \mathbf{\Delta}(t)\right)$ is given by $\mathbf{B} \mathbf{D_A} \operatorname{diag}(\dot{f}(\mathbf{B}^T \mathbf{\Delta})) \mathbf{B}^T$, which is positive definite for all the vectors $\mathbf{\Delta}$ that are equilibria of (66). Denoting by $\mathbf{\Delta}^\star \triangleq [\Delta_1^\star, \ldots, \Delta_N^\star]^T$ one of the (isolated) equilibria of (66), i.e. satisfying

$$\Delta \boldsymbol{\omega} = K \mathbf{D_c}^{-1} \mathbf{B} \mathbf{D_A} f\left(\mathbf{B}^T \mathbf{\Delta}^\star\right), \quad (67)$$

by translating the origin to the equilibrium $\mathbf{\Delta}^\star$, we can make $\mathbf{0}_N$ to be an equilibrium of (66) and write[22]

$$\dot{\mathbf{\Delta}}(t) = -K \mathbf{D_c}^{-1} \mathbf{B} \mathbf{D_A} \left[f\left(\mathbf{B}^T \mathbf{\Delta}(t) + \mathbf{B}^T \mathbf{\Delta}^\star\right) - f\left(\mathbf{B}^T \mathbf{\Delta}^\star\right)\right]. \quad (68)$$

To make explicit the dependence of (68) on the weighted Laplacian, we introduce the following function

$$g_{\Delta_j^\star, \Delta_i^\star}(x) \triangleq f\left(x + \Delta_j^\star - \Delta_i^\star\right) - f\left(\Delta_j^\star - \Delta_i^\star\right), \quad (69)$$

that, because of **a1**, satisfies the following properties

$$g_{\Delta_j^\star, \Delta_i^\star}(x) = -g_{\Delta_i^\star, \Delta_j^\star}(-x),$$

$$\frac{g_{\Delta_j^\star, \Delta_i^\star}(x)}{x} = \begin{cases} \left.\dfrac{df(x)}{dx}\right|_{x=\Delta_j^\star - \Delta_i^\star} > 0, & x = 0, \\ > 0, & x \neq 0, \end{cases} \quad \forall \mathbf{\Delta}^\star : \Delta_j^\star \neq \Delta_i^\star. \quad (70)$$

---
[21] An equilibrium point is isolated if it has a surrounding neighborhood containing no other equilibria.

[22] With a slight abuse of notation, we use the same variables $\mathbf{\Delta}$, to denote also the system (66), after the shift around $\mathbf{\Delta}^\star$.

Using (69) and introducing the diagonal matrix $\mathbf{D}_{\boldsymbol{\Delta}^\star,\boldsymbol{\Delta}}$, whose positive (see (70)) diagonal entries are given by $g_{\Delta_j^\star,\Delta_i^\star}(\Delta_j(t) - \Delta_i(t))/(\Delta_j(t) - \Delta_i(t))$, indexed from 1 to $E$, and $g_{\Delta_j^\star,\Delta_i^\star}(\cdot)$ is defined in (69), system (68) can be equivalently rewritten as

$$\dot{\boldsymbol{\Delta}}(t) = -K\mathbf{D}_\mathbf{c}^{-1}\mathbf{B}\mathbf{D}_\mathbf{A}\mathbf{D}_{\boldsymbol{\Delta}^\star,\boldsymbol{\Delta}}\mathbf{B}^T\boldsymbol{\Delta}(t)$$
$$\triangleq -K\mathbf{D}_\mathbf{c}^{-1}\mathbf{L}_{\mathbf{A},\boldsymbol{\Delta}^\star,\boldsymbol{\Delta}}\boldsymbol{\Delta}(t), \tag{71}$$

where $\mathbf{L}_{\mathbf{A},\boldsymbol{\Delta}^\star,\boldsymbol{\Delta}}$ is the weighted Laplacian of the graph, with diagonal weigth-matrix $\mathbf{D}_\mathbf{A}\mathbf{D}_{\boldsymbol{\Delta}^\star,\boldsymbol{\Delta}}$.

The system (71) admits the point $\mathbf{0}_N$ as the unique equilibrium (which also guarantees the uniqueness of $\boldsymbol{\Delta}^\star$ for (66)) and such an equilibrium is globally asymptotically stable, as we argue next.

From **a1** and the properties of the weighed Laplacian of a connected graph, we have that

$$\mathbf{L}_{\mathbf{A},\boldsymbol{\Delta}^\star,\boldsymbol{\Delta}}\boldsymbol{\Delta} = \mathbf{0}_N \Leftrightarrow \boldsymbol{\Delta} \in \text{span}\{\mathbf{1}_N\}, \quad \forall \boldsymbol{\Delta}^\star. \tag{72}$$

From (65) and (72), it follow that all the equilibria $\overline{\boldsymbol{\Delta}}$ of system (71) are given by

$$\overline{\boldsymbol{\Delta}} \in \{\text{span}\{\mathbf{1}_N\} \cap \text{span}\{\mathbf{c}_\perp\}\} = \{\mathbf{0}_N\}, \tag{73}$$

where $\text{span}\{\mathbf{c}_\perp\} \triangleq \{\mathbf{x} \in \mathbb{R}^N : \mathbf{1}_N^T\mathbf{D}_\mathbf{c}\mathbf{x} = 0\}$. Hence, the *unique* equilibrium of (71) is the vector $\overline{\boldsymbol{\Delta}} = \mathbf{0}_N$.

After having showed the uniqueness of the equilibrium, we can prove its global asymptotic stability. To this end, consider the following candidate positive definite[23] Lyapunov function

$$V(\boldsymbol{\Delta}(t)) = \frac{1}{2} \| \mathbf{D}_\mathbf{c}^{1/2}\boldsymbol{\Delta}(t) \|^2. \tag{74}$$

The function $V(\boldsymbol{\Delta}(t))$ in (74) is non-increasing along trajectories of (71), since

$$\dot{V}(\boldsymbol{\Delta}(t)) = \boldsymbol{\Delta}^T(t)\mathbf{D}_\mathbf{c}\dot{\boldsymbol{\Delta}}(t)$$
$$= -K\boldsymbol{\Delta}^T(t)\mathbf{L}_{\mathbf{A},\boldsymbol{\Delta}^\star,\boldsymbol{\Delta}(t)}\boldsymbol{\Delta}(t) \leq 0, \tag{75}$$

where the last inequality follows from $\mathbf{L}_{\mathbf{A},\boldsymbol{\Delta}^\star,\boldsymbol{\Delta}(t)} \succeq \mathbf{0}_{N\times N}, \forall \boldsymbol{\Delta}^\star, \boldsymbol{\Delta}(t) \in \mathbb{R}^n$, and equality in (75) is reached if and only if $\mathbf{L}_{\mathbf{A},\boldsymbol{\Delta}^\star,\boldsymbol{\Delta}}\boldsymbol{\Delta} = \mathbf{0}_N$. Since, by definition, $\boldsymbol{\Delta}$ must satisfy also the constraint $\mathbf{1}_N^T\mathbf{D}_\mathbf{c}\boldsymbol{\Delta} = 0$ (see (65)), the function $\dot{V}(\boldsymbol{\Delta}) = 0$ if and only if (73) holds true, i.e. $\boldsymbol{\Delta} = \mathbf{0}_N$. Hence, $V(\boldsymbol{\Delta})$ in (74) is a valid Lyapunov function for (71) and $\boldsymbol{\Delta} = \mathbf{0}_N$ is globally asymptotically stable for (71) [42, Theorem 5.24]. This guarantees that the state $\boldsymbol{\Psi}(t)$ of (34) converges to an equilibrium of (34) as $t \to +\infty$, for any set of initial conditions.

Observe that, through the whole proof, we have always considered positive values of $K$. This comes from the fact that, all the potential equilibria of the system (34) (and thus of (71)) corresponding to negative $K$, are instable. In fact, for negative $K$, the valid Lyapunov function $V(\mathbf{\Delta}(t))$ defined in (74), has first derivative $\dot{V}(\mathbf{\Delta}(t))$ along trajectories of (71) that is positive definite, which proves the instability of the equilibrium $\mathbf{\Delta} = \mathbf{0}_N$ [42, Theorem 5.29].

---

[23]A continuous function $V : \mathbb{R}^n \mapsto \mathbb{R}$ is called a *positive definite function* if $V(\mathbf{0}) = 0$, and $V(\mathbf{x}) \geq \alpha(|\mathbf{x}|), \forall \mathbf{x} \in \mathbb{R}^n$, where $\alpha : \mathbb{R} \mapsto \mathbb{R}$ is some continuous, strictly increasing scalar function, with $\alpha(0) = 0$, and $\alpha(p) \mapsto \infty$ as $p \mapsto \infty$ [42, Definition 5.13].

## B Proof of Theorem 2

Since most of the proof of Theorem follows the same approach, already described in the Appendix A, in the following we point out only the differences.

### B.1 Existence of the synchronized state.

Given the system (5), we introduce, as for (34), the following change of variables

$$\mathbf{\Psi}_i(t) = \boldsymbol{\theta}_i(t) - \boldsymbol{\omega}_L^\star t, \quad i = 1, 2, \ldots, N, \tag{76}$$

with $\boldsymbol{\omega}_L^\star$ defined in (11), and rewrite (5) as

$$\dot{\mathbf{\Psi}}(t) = \Delta \boldsymbol{\omega} - K \mathbf{D}_\mathbf{Q}^{-1} \mathbf{P}^T (\mathbf{I}_L \otimes \mathbf{B} \mathbf{D}_\mathbf{A}) f\left((\mathbf{I}_L \otimes \mathbf{B}^T) \mathbf{P} \mathbf{\Psi}(t)\right), \tag{77}$$

where $\mathbf{\Psi}(t) \triangleq [\mathbf{\Psi}_1(t), \ldots, \mathbf{\Psi}_N(t)]^T$, with $\mathbf{\Psi}_i(0) = \boldsymbol{\theta}_i(0)$, and $\Delta \boldsymbol{\omega} = \boldsymbol{\omega} - (\mathbf{1}_N \otimes \boldsymbol{\omega}_L^\star)$. To obtain (77), we have used the following chain of equalities $\mathbf{P}(\mathbf{1}_N \otimes \boldsymbol{\omega}_L^\star) = (\boldsymbol{\omega}_L^\star \otimes \mathbf{1}_N)$, and $(\mathbf{I}_L \otimes \mathbf{B}^T)(\boldsymbol{\omega}_L^\star \otimes \mathbf{1}_N) = \text{diag}(\boldsymbol{\omega}_L^\star) \otimes \mathbf{B}^T \mathbf{1}_N = \mathbf{0}_{LE \times L}$. Given (77), the synchronized state of (5) exists if and only if the following system of non-linear equations admits a solution

$$\mathbf{PD}_\mathbf{Q} \Delta \boldsymbol{\omega} = K(\mathbf{I}_L \otimes \mathbf{B} \mathbf{D}_\mathbf{A}) f\left((\mathbf{I}_L \otimes \mathbf{B}^T) \mathbf{P} \mathbf{\Psi}\right). \tag{78}$$

We recast now the study of the existence of a solution for (78), to the study of the solution of a set of simpler sub-systems, similar to (42). To this end, we introduce the vectors $\bar{\mathbf{\Psi}} \triangleq \mathbf{P}\mathbf{\Psi}$ and $\Delta \bar{\boldsymbol{\omega}} \triangleq \mathbf{PD}_\mathbf{Q} \Delta \boldsymbol{\omega}$, partitioned as $\bar{\mathbf{\Psi}} = [\bar{\mathbf{\Psi}}_1^T, \ldots, \bar{\mathbf{\Psi}}_L^T]^T$, and $\Delta \bar{\boldsymbol{\omega}} = [\Delta \bar{\boldsymbol{\omega}}_1^T, \ldots, \Delta \bar{\boldsymbol{\omega}}_L^T]^T$, with $\bar{\mathbf{\Psi}}_i \triangleq [\Psi_1^{(i)}, \ldots, \Psi_N^{(i)}]^T$ and $\Delta \bar{\boldsymbol{\omega}}_i \triangleq [\Delta \omega_1^{(i)}, \ldots, \Delta \omega_N^{(i)}]^T$, $i = 1, \ldots, L$, where $\Psi_j^{(i)}$ denotes the $i$-th component of the $j$-th sensor's state vector $\mathbf{\Psi}_j$, and $\Delta \omega_j^{(i)}$ is the $i$-th component of the vector $\boldsymbol{\omega}_j - \boldsymbol{\omega}_L^\star$. Then, the system (78) can be equivalently rewritten as

$$\mathbf{L}_{\bar{\mathbf{\Psi}}_i, \mathbf{A}} \bar{\mathbf{\Psi}}_i = \frac{1}{K} \Delta \bar{\boldsymbol{\omega}}_i, \quad i = 1, \ldots, L, \tag{79}$$

where $\mathbf{L}_{\bar{\mathbf{\Psi}}_i, \mathbf{A}} \triangleq \mathbf{B} \mathbf{D}_\mathbf{A} \mathbf{D}_{\bar{\mathbf{\Psi}}_i} \mathbf{B}^T$ is the weighted Laplacian of the graph, with diagonal weights-matrix $\mathbf{D}_\mathbf{A} \mathbf{D}_{\bar{\mathbf{\Psi}}_i}$, and $\mathbf{D}_{\bar{\mathbf{\Psi}}_i}$ is still given by (37), by replacing $\mathbf{\Psi}$ with $\bar{\mathbf{\Psi}}_i$. Thus, a solution of (78) exists if and only if every equation in (79) admits a fixed point in $\mathbb{R}^N$. But, for any given $i$, (79) is equivalent to (42), if $\mathbf{L}_{\mathbf{A}, \mathbf{\Psi}}$ and $\mathbf{D}_\mathbf{c} \Delta \boldsymbol{\omega}$ are replaced with $\mathbf{L}_{\bar{\mathbf{\Psi}}_i, \mathbf{A}}$ and $\Delta \bar{\boldsymbol{\omega}}_i$, respectively. Hence, each equation in (79) admits a fixed point provided that $K > K_U^{(i)}$, whereas a solution of (79) cannot exist if $K < K_L^{(i)}$. A lower bound for $K_L^{(i)}$ and an upper bound for $K_U^{(i)}$ are given by

$$K_L^{(i)} \geq \frac{\|\Delta\bar{\boldsymbol{\omega}}_i\|_\infty}{f_{\max}d_{\max}},, \quad \text{and} \quad K_U^{(i)} \leq \frac{\|\Delta\bar{\boldsymbol{\omega}}_i\|_2}{g\lambda_2(\mathbf{L_A})}, \tag{80}$$

with $g$ defined in (18).

Since the coupling among the equations in (79) is given only by the presence of $K$, a fixed point for all the equations exists if and only if the critical values $K_L$ and $K_U$ in (79) are chosen as the maximum of $\{K_L^{(i)}\}_i$ and $\{K_U^{(i)}\}_i$ respectively, with $K_L^{(i)}$ and $K_U^{(i)}$ defined in (80). This proves (20).

The upper bound of $K_U$ and lower bounds of $K_L$ in (21) can be obtained following the same approach used in Appendix A.1 to prove (19).

## B.2 Global stability of the synchronized state

We consider the system (77) and show that, under **a1** − **a3** and $K > K_U$, the state vector converges to an equilibrium, from any set of initial conditions, which proves the globally asymptotic stability of the synchronized state of (5). Since, similarly to (63)

$$(\mathbf{1}_N^T \otimes \mathbf{I}_L)\mathbf{D_Q}\dot{\boldsymbol{\Psi}}(t) = \sum_{i=1}^N \mathbf{Q}_i \dot{\boldsymbol{\Psi}}_i(t) = \mathbf{0}_L \quad \Leftrightarrow \quad \sum_{i=1}^N \mathbf{Q}_i \boldsymbol{\Psi}_i(t) = \sum_{i=1}^N \mathbf{Q}_i \boldsymbol{\Psi}_i(0), \quad \forall t \geq 0, \tag{81}$$

$(\mathbf{1}_N^T \otimes \mathbf{I}_L)\mathbf{D_Q}\dot{\boldsymbol{\Psi}}(t)$ is an invariant and thus we can decompose $\boldsymbol{\Psi}(t)$ in (77) according to

$$\boldsymbol{\Psi}(t) \triangleq (\mathbf{1}_{N\times N} \otimes \mathbf{I}_L)\overline{\mathbf{D}}_{\mathbf{Q}}\boldsymbol{\Psi}(0) + \boldsymbol{\Delta}(t), \tag{82}$$

where $\mathbf{1}_{N\times N}$ is the $N\times N$ matrix of all ones, $\overline{\mathbf{D}}_{\mathbf{Q}} \triangleq \left(\mathbf{I}_N \otimes \left(\sum_{i=1}^N \mathbf{Q}_i\right)^{-1}\right)\mathbf{D_Q}$, and $\boldsymbol{\Delta}(t) \triangleq [\boldsymbol{\Delta}_1^T(t),\ldots,\boldsymbol{\Delta}_N^T(t)]^T$ satisfies the constraint

$$(\mathbf{1}_N^T \otimes \mathbf{I}_L)\mathbf{D_Q}\boldsymbol{\Delta}(t) = \sum_{i=1}^N \mathbf{Q}_i \boldsymbol{\Delta}_i(t) = \mathbf{0}_L. \tag{83}$$

Introducing (82) in (77), we obtain the following dynamic for $\boldsymbol{\Delta}(t)$:

$$\dot{\boldsymbol{\Delta}}(t) = \Delta\boldsymbol{\omega} - K\mathbf{D_Q}^{-1}\mathbf{P}^T(\boldsymbol{I}_L \otimes \boldsymbol{B}\boldsymbol{D_A})f\left((\boldsymbol{I}_L \otimes \boldsymbol{B}^T)\mathbf{P}\boldsymbol{\Delta}(t)\right), \tag{84}$$

where we used the chain of equalities $(\boldsymbol{I}_L \otimes \boldsymbol{B}^T)\mathbf{P}(\mathbf{1}_{N\times N} \otimes \mathbf{I}_L) = (\boldsymbol{I}_L \otimes \boldsymbol{B}^T)(\mathbf{I}_L \otimes \mathbf{1}_N)(\mathbf{1}_N^T \otimes \mathbf{I}_L) = (\boldsymbol{I}_L \otimes \boldsymbol{B}^T\mathbf{1}_N)(\mathbf{1}_N^T \otimes \mathbf{I}_L) = \mathbf{0}_{LE\times LN}$.

Following the same approach used in Appendix A.2 to obtain (68), we can translate the system (84) around the isolated equilibrium $\boldsymbol{\Delta}^\star$ given by

$$\begin{cases} \Delta\boldsymbol{\omega} = K\mathbf{D_Q}^{-1}\mathbf{P}^T(\boldsymbol{I}_L \otimes \boldsymbol{B}\boldsymbol{D_A})f\left((\boldsymbol{I}_L \otimes \boldsymbol{B}^T)\mathbf{P}\boldsymbol{\Delta}^\star\right), \\ (\mathbf{1}_N^T \otimes \mathbf{I}_L)\mathbf{D_Q}\boldsymbol{\Delta}^\star = \mathbf{0}_L, \end{cases} \tag{85}$$

so that the vector $\mathbf{0}_{LN}$ is an isolated equilibrium of the following translated system[24]

$$\dot{\boldsymbol{\Delta}}(t) = -K\mathbf{D_Q}^{-1}\mathbf{P}^T(\boldsymbol{I}_L \otimes \boldsymbol{B}\boldsymbol{D_A})\left(f\left((\boldsymbol{I}_L \otimes \boldsymbol{B}^T)\mathbf{P}(\boldsymbol{\Delta}(t) + \boldsymbol{\Delta}^\star)\right) - f\left((\boldsymbol{I}_L \otimes \boldsymbol{B}^T)\mathbf{P}\boldsymbol{\Delta}^\star\right)\right). \tag{86}$$

---
[24]We use the same variables to denote the translated system (84) around $\boldsymbol{\Delta}^\star$.

We rewrite now the system (86) in a more compact form, making explicit the dependence on the weighted Laplacian. To this end, we introduce the vectors $\bar{\boldsymbol{\Delta}}(t) \triangleq \mathbf{P}\boldsymbol{\Delta}(t)$ and $\bar{\boldsymbol{\Delta}}^\star \triangleq \mathbf{P}\boldsymbol{\Delta}^\star$, partitioned as $\bar{\boldsymbol{\Delta}}(t) = [\bar{\boldsymbol{\Delta}}_1^T(t), \ldots, \bar{\boldsymbol{\Delta}}_L^T(t)]^T$, and $\bar{\boldsymbol{\Delta}}^\star = [\bar{\boldsymbol{\Delta}}_1^{\star T}, \ldots, \bar{\boldsymbol{\Delta}}_L^{\star T}]^T$, with $\bar{\boldsymbol{\Delta}}_i(t) \triangleq [\Delta_1^{(i)}(t), \ldots, \Delta_N^{(i)}(t)]^T$ and $\bar{\boldsymbol{\Delta}}_i^\star \triangleq [\Delta_1^{\star(i)}, \ldots, \Delta_N^{\star(i)}]^T$, where $\Delta_j^{(i)}(t)$ and $\Delta_j^{\star(i)}$ denote the $i$-th component of $\boldsymbol{\Delta}_j(t)$ and $\boldsymbol{\Delta}_j^\star$, respectively. Then, the system (86) can be equivalently rewritten as

$$\dot{\boldsymbol{\Delta}}(t) = -K\mathbf{D}_{\mathbf{Q}}^{-1}\mathbf{P}^T \mathbf{L}_{\mathbf{A},\boldsymbol{\Delta}^\star,\boldsymbol{\Delta}} \mathbf{P}\boldsymbol{\Delta}(t), \tag{87}$$

where $\boldsymbol{\Delta}(t)$ satisfies the constraint (83), and $\mathbf{L}_{\mathbf{A},\boldsymbol{\Delta}^\star,\boldsymbol{\Delta}} \triangleq (\boldsymbol{I}_L \otimes \boldsymbol{B})\mathbf{D}_{\mathbf{A},\boldsymbol{\Delta}^\star,\boldsymbol{\Delta}}(\boldsymbol{I}_L \otimes \boldsymbol{B}^T)$, with $\mathbf{D}_{\mathbf{A},\boldsymbol{\Delta}^\star,\boldsymbol{\Delta}} \triangleq (\boldsymbol{I}_L \otimes \mathbf{D}_A) \operatorname{diag}(\mathbf{D}_{\bar{\boldsymbol{\Delta}}_1^\star,\bar{\boldsymbol{\Delta}}_1}, \ldots, \mathbf{D}_{\bar{\boldsymbol{\Delta}}_L^\star,\bar{\boldsymbol{\Delta}}_L})$ and $\mathbf{D}_{\bar{\boldsymbol{\Delta}}_k^\star,\bar{\boldsymbol{\Delta}}_k}$ is the $E \times E$ diagonal positive matrix, whose diagonal entries are given by $g_{\Delta_i^{\star(k)},\Delta_j^{\star(k)}}(\Delta_j^{(k)} - \Delta_i^{(k)})/(\Delta_j^{(k)} - \Delta_i^{(k)})$ indexed from 1 to $E$, and $g_{\Delta_i^{\star(k)},\Delta_j^{\star(k)}}(\cdot)$ is the same function as defined in (69), with $\Delta_i$ and $\Delta_i^\star$ replaced by $\Delta_i^{(k)}$ and $\Delta_j^{\star(k)}$, respectively.

Using the same technique as in Appendix A.2 one can prove that the vector $\mathbf{0}_{LN}$ is the unique equilibrium of (87) and it is globally asymptotically stable. A valid positive definite Lyapunov function for (87) is

$$V(\boldsymbol{\Delta}) = 1/2 \parallel \mathbf{D}_{\mathbf{Q}}^{1/2}\boldsymbol{\Delta}(t) \parallel^2, \tag{88}$$

that can be seen to be non-increasing along trajectories of (86), and with zero derivative if and only $\boldsymbol{\Delta} = \mathbf{0}_{LN}$. This proves the globally asymptotically stability of the synchronized state of (5).